\documentclass[10pt,journal,compsoc]{IEEEtran}
\ifCLASSOPTIONcompsoc
  \usepackage[nocompress]{cite}
\else
  \usepackage{cite}
\fi
\ifCLASSINFOpdf
   \usepackage[pdftex]{graphicx}
   \graphicspath{{../pdf/}{../jpeg/}}
   \DeclareGraphicsExtensions{.pdf,.jpeg,.png}
\else
   \usepackage[dvips]{graphicx}
   \graphicspath{{../eps/}}
   \DeclareGraphicsExtensions{.eps}
\fi
\usepackage{url}
\usepackage{amsmath}
\ifCLASSINFOpdf
 \usepackage[pdftex]{thumbpdf}
\else
 \usepackage[dvips]{thumbpdf}
\fi

\usepackage{enumitem}
\usepackage{multirow}
\usepackage{amssymb}

\usepackage{subfigure,epsfig}
\usepackage{algorithm}
\usepackage{algorithmic}
\usepackage{multirow}
\usepackage{float}
\usepackage{url}
\usepackage{amsmath}
\renewcommand{\algorithmicrequire}{\textbf{Input:}} 
\renewcommand{\algorithmicensure}{\textbf{Output:}} 

\usepackage{multirow}
\usepackage{enumitem}

\usepackage{array}
\usepackage{booktabs}
\usepackage{color}
\renewcommand\arraystretch{1}
\setlist[itemize]{leftmargin=*}
\setlist[enumerate]{leftmargin=*}

\newcommand{\minitab}[2][l]{\begin{tabular}{#1}#2\end{tabular}}

\def\BibTeX{{\rm B\kern-.05em{\sc i\kern-.025em b}\kern-.08emT\kern-.1667em\lower.7ex\hbox{E}\kern-.125emX}}
    
\begin{document}

\title{Causal Inference for Knowledge Graph based Recommendation}

\author{Yinwei~Wei,~\IEEEmembership{Member,~IEEE,}
        Xiang~Wang,~\IEEEmembership{Member,~IEEE,}
        Liqiang~Nie,~\IEEEmembership{Senior~Member,~IEEE,}\\
        Shaoyu~Li,
        Dingxian~Wang,
        Tat-Seng~Chua~\IEEEmembership{Member,~IEEE}
\IEEEcompsocitemizethanks{
\IEEEcompsocthanksitem Yinwei Wei and Tat-Seng Chua are with School of Computing, National University of Singapore, Singapore. \protect(E-mail: weiyinwei@hotmail.com; chuats@comp.nus.edu.sg)
\IEEEcompsocthanksitem Xiang Wang is with School of Information Science and Technology, University of Science and Technology of China, China. \protect(E-mail: xiangwang1223@gmail.com)
\IEEEcompsocthanksitem Liqiang Nie and Shaoyu Li are with Shandong University, China. \protect(E-mail: nieliqiang@gmail.com; lssy97@163.com)
\IEEEcompsocthanksitem D. Wang is with Ranking team, Search Science Department, eBay Inc,
China. \protect(E-mail: diwang@ebay.com)
\IEEEcompsocthanksitem Liqiang Nie is the corresponding author.}
}



\markboth{IEEE Transactions on Knowledge and Data Engineering}%
{Shell \MakeLowercase{\textit{et al.}}: Bare Advanced Demo of IEEEtran.cls for IEEE Computer Society Journals}
\IEEEtitleabstractindextext{%

\begin{abstract}


Knowledge Graph~(KG), as a side-information, tends to be utilized to supplement the collaborative filtering~(CF) based recommendation model. By mapping items with the entities in KGs, prior studies mostly extract the knowledge information from the KGs and inject it into the representations of users and items. 
Despite their remarkable performance, they fail to model the user preference on attribute in the KG, since they ignore that (1) the structure information of KG may hinder the user preference learning, and (2) the user's interacted attributes will result in the bias issue on the similarity scores.

With the help of causality tools, we construct the causal-effect relation between the variables in KG-based recommendation and identify the reasons causing the mentioned challenges. 
Accordingly, we develop a new framework, termed Knowledge Graph-based Causal Recommendation~(KGCR), which implements the deconfounded user preference learning and adopts counterfactual inference to eliminate bias in the similarity scoring. Ultimately, we evaluate our proposed model on three datasets, including Amazon-book, LastFM, and Yelp2018 datasets. By conducting extensive experiments on the datasets, we demonstrate that KGCR outperforms several state-of-the-art baselines, such as KGNN-LS~\cite{KGNN-LS}, KGAT~\cite{KGAT} and KGIN~\cite{KGIN}. We release our codes in the public domain: https://github.com/weiyinwei/KGCR.
\end{abstract}
}

\maketitle

\section{Introduction}
Knowledge graph~(KG), as an external knowledge, has been witnessed in improving the collaborative filtering~(CF) based recommender system. 
Typically, KG is a directed graph consisting of real-world facts, where the nodes function as the entities and the edges reflect the relations between the entities~\cite{KG}. By mapping the items to entities in the KG, it is able to offer the external knowledge to CF-based models~\cite{survey_new}. 
For the sake of description, we term the entities corresponding to items as item entities and the other entities for depicting the items in KGs as the attribute entities.

By investigating the existing KG-based recommendation models, we roughly divide the prior studies into three groups~\cite{Survey}: 
Embedding-based models which conduct the knowledge graph embedding (KGE) algorithms on KGs, and then enrich the CF-based user and item representations by incorporating the embeddings of item entities~\cite{CKE,DKN}; 
Path-based models that leverage the similarity of connectivity patterns between the item entities in KGs, in order to provide the external cues to enhance the collaborative signal~\cite{Pathsim,SI2}; 
and GCN-based models which explicitly pass the knowledge information to the item and user embeddings upon user-item graphs and KGs in an end-to-end manner~\cite{KGAT,KGIN}. 
Accordingly, we can find that they mostly utilize the knowledge information in KGs through the item entity and propagate it along the user-item interactions.


\begin{figure}
	\centering
    \includegraphics[width=0.35\textwidth]{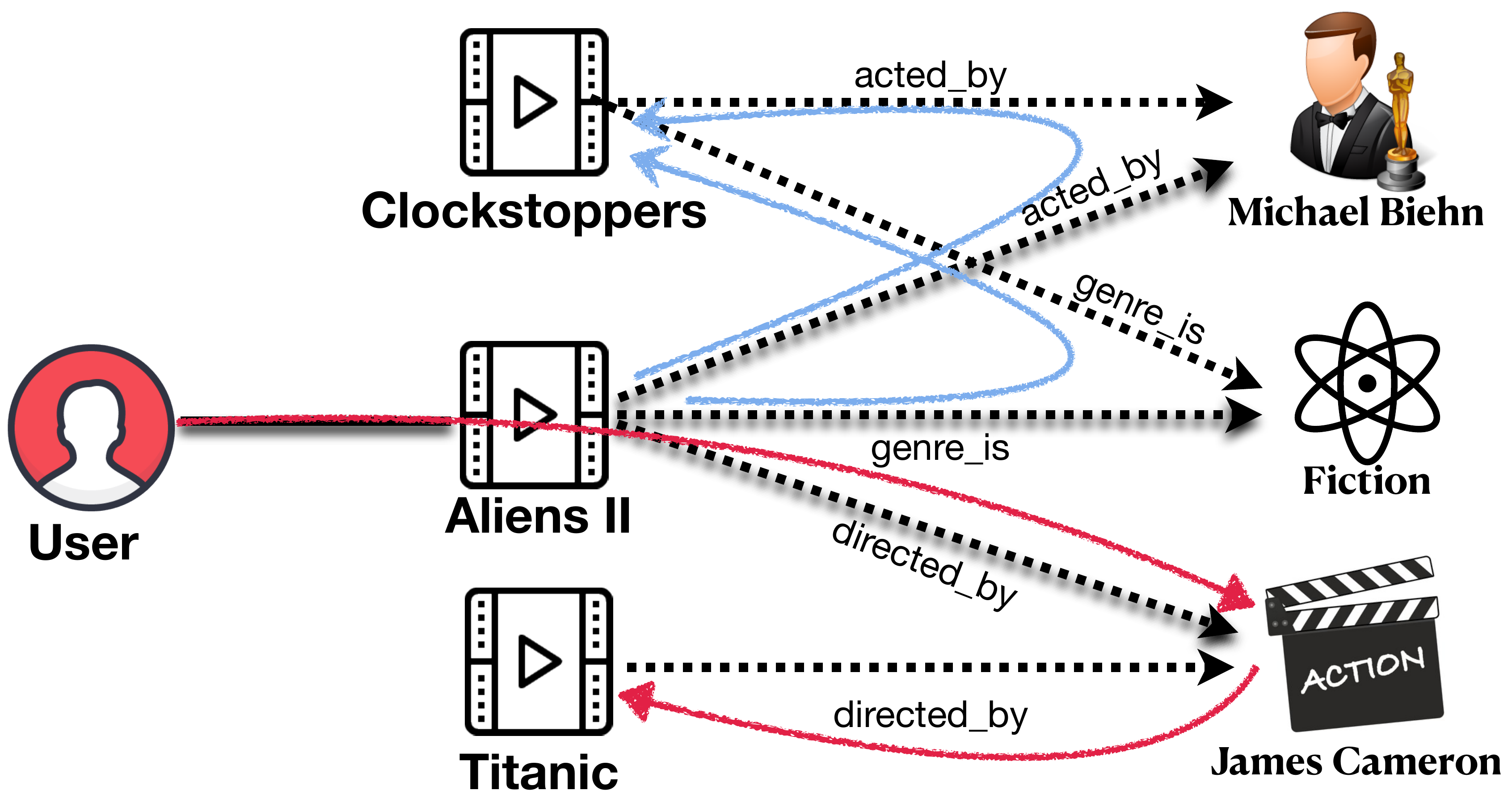}
    \caption{An illustration of KG-based recommendation. }
    \vspace{-15pt}
	\label{fig:figure1}
\end{figure}
Despite their promising results, we argue that these methods fail to model the fine-grained user preference on attribute entities in the KG. In the real scenario, people are prone to select the items for some particular attributes, \textit{e.g.,} someone may watch a movie for its director or buy shoes due to its brand. As illustrated in Figure~\ref{fig:figure1}, \textit{User}'s preference on the attributes \textit{Fiction}, \textit{Michael Biehn}, and \textit{James Cameron} which she/he interacted through the movie \textit{Aliens II} heavily affects the recommendation of $Clockstoppers$ and $Titanic$. 
It seems that the exiting models, such as KGAT~\cite{KGAT} and KGIN~\cite{KGIN}, can learn the fine-grained user preference from her/his interacted attributes~(short for the attributes that user interacted through items) and score the similarity between the user and item representations~\cite{KVMN}. However, they suffer from the following two challenges:
\begin{itemize}
    \item The structure information of KGs hinders the representation of user preference on attribute entities. This is because the information decides which attributes associate with items and how the attributes~(\textit{e.g.,} \textit{Fiction}, \textit{Michael Biehn}, and \textit{James Cameron}) are entangled by the items~(\textit{e.g.,} \textit{Aliens II}). 
    Merely depending on the user-item interactions, it is hard to distinguish the user preferences to these different attributes, inevitably affecting the user preference learning on attributes. 
    \textit{Therefore, how to utilize the user-item interactions to learn the user preference on attribute is the first challenge we are facing.}
    \item The user's interacted attributes will result in the bias issue on the similarity scores, which may offer inaccurate supervision signal during the training phase. 
    Specifically, the items with more attributes exposed in users' interacted attributes are more likely to achieve the higher scores~\cite{DecRS}. In the same sample shown in Figure~\ref{fig:figure1}, the \textit{Fiction} movie \textit{Clockstoppers} acted by \textit{Michael Biehn} may score higher than \textit{Titanic} directed by \textit{James Cameron}. Because the attributes provide more shortcuts~(denoted as blue curves) to \textit{Clockstoppers}, which bypasses the user preference to \textit{James Cameron}~(denoted as red curve). \textit{Thus, how to eliminate the bias from the similarity scores is the other key challenge.}
\end{itemize}

\begin{figure}
	\centering
	\subfigure[Proposed Causal Graph]{
    \includegraphics[width=0.2\textwidth]{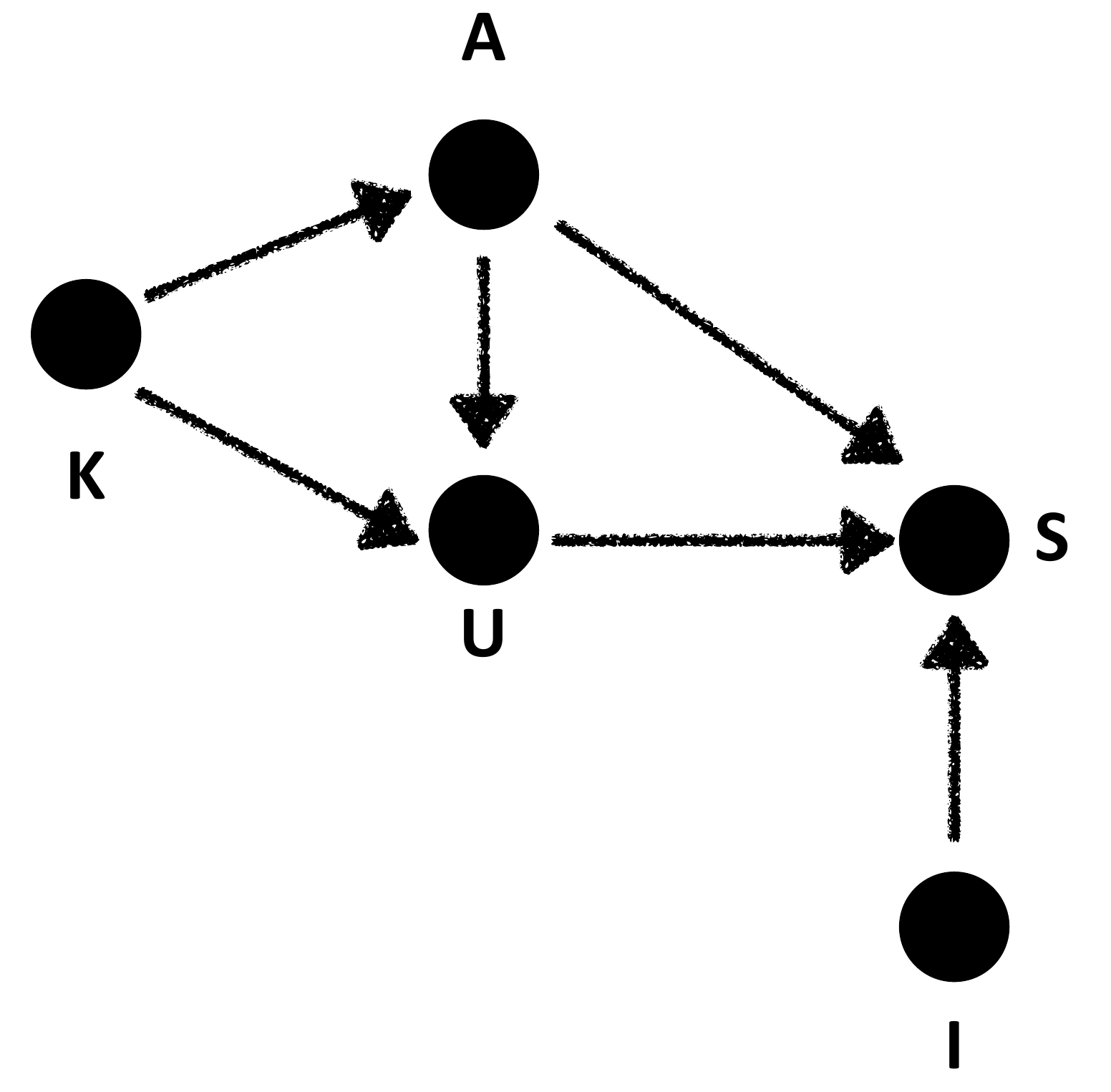}
	\label{fig:figure2_1}
    }
    \subfigure[Intervention on Causal Graph]{
    \includegraphics[width=0.2\textwidth]{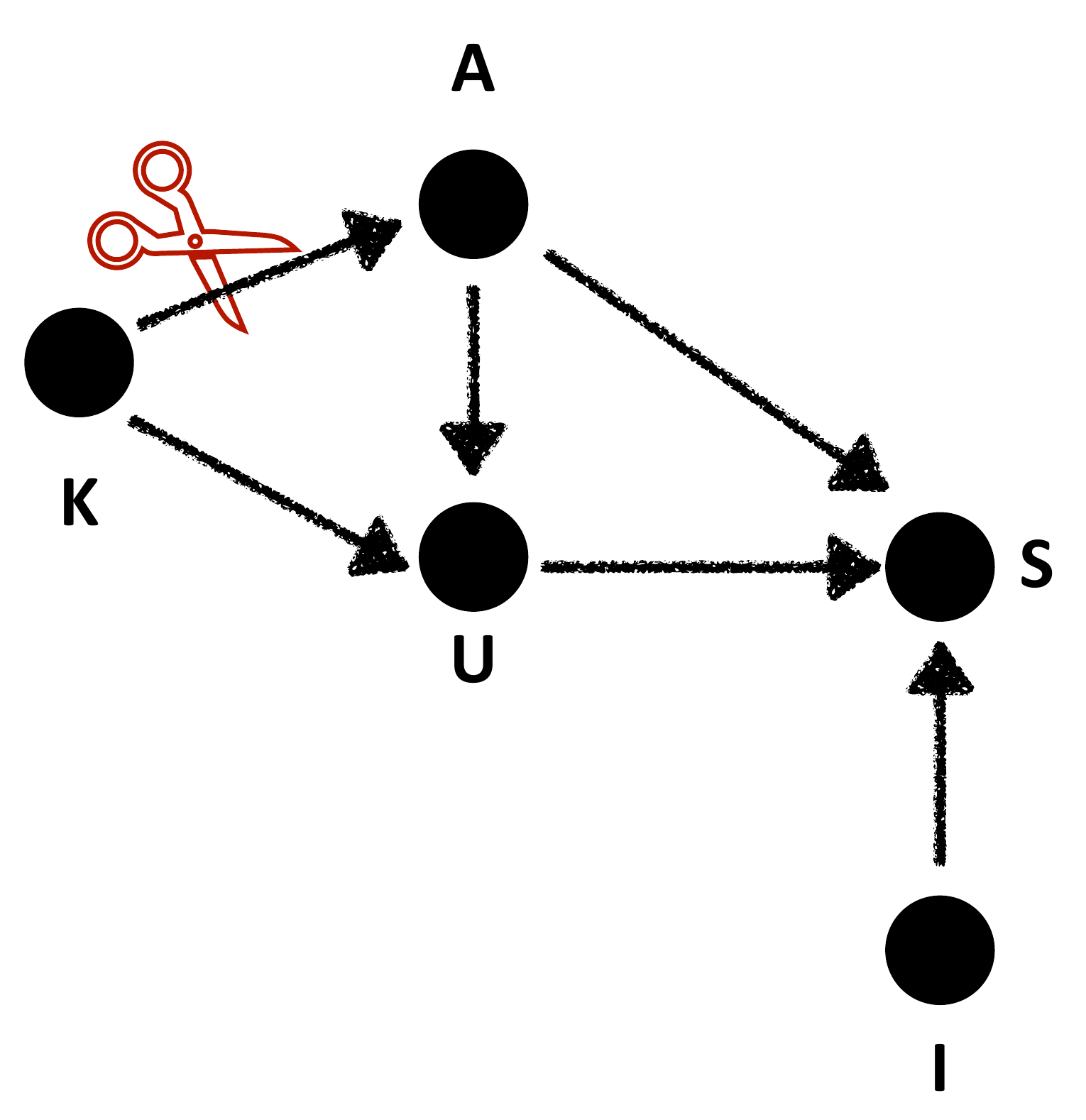}
	\label{fig:figure2_2}
    }
\vspace{-5pt}
    \caption{Illustration of our constructed causal graph.}
\vspace{-5pt}
\end{figure}

To address the challenges, we resort to the language of causal inference~\cite{causality} to explore the reasons causing the limitations. 
Specifically, we construct a causal graph to model the causal-effect factors of the similarity scoring, as shown in Figure~\ref{fig:figure2_1}. 
We first consider how the structure information~($K$) affects the user preference on attribute~($U$). In the causal graph, $K \rightarrow A$ illustrates that the structure information decides the user's interacted attributes $A$, and $K \rightarrow U$ reflects the entanglement of attributes observed by user. 
Obviously, $K$, as a confounder, opens up a back-door path $U \leftarrow K \rightarrow A$, giving rise to the spurious relation between the user preference and interacted attributes. To reduce its negative influence, we conduct the intervention operation on the causal graph and devise a Knowledge Graph-based Causal Recommendation~(KGCR) model, which implements the deconfounded user preference learning and models the user-item similarity score. 
As for the bias issue in similarity scoring, we attribute it to the direct effect of $A$ on $S$. The interacted attribute~($A$) provides a shortcut for the prediction of user-item interaction, bypassing the user preference on attribute $U$. Hence, we adopt the counterfactual inference to eliminate the bias caused by $A \rightarrow S$. In particular, we estimate \textit{total effects}~(TE) on the similarity score in the factual world and \textit{natural direct effect}~(NDE) of the interacted attributes on the score in the counterfactual world. Thereafter, we can remove NDE from TE during the inference phase to answer the question: \textit{what the score would be, if we only obtain the user and item representations based on the attributes}. 

To evaluate our proposed model, we conduct extensive experiments on three publicly accessible datasets, including Amazon-book, LastFM, and Yelp2018 datasets. The results show that our proposed model outperforms several state-of-the-art baselines, such as KGNN-LS~\cite{KGNN-LS}, KGAT~\cite{KGAT} and KGIN~\cite{KGIN}. Furthermore, we do ablation studies to verify the effectiveness and rationality of our constructed causal graph and implementations in KGCR. 

In a nutshell, the main contributions of this work are threefold:
\begin{itemize}
    \item By investigating the existing KG-based recommendation models, we identify two key challenges in modeling the user preference on attribute in KGs. 
    \item We leverage the causality techniques to explore the causal-effect relation between the variables in the KG-based recommendation and discover the potential issues challenging us. To the best of our knowledge, this is the first attempt to analyze and optimize the KG-based recommendation through the lens of causality.
    \item We develop Knowledge Graph-based Causal Recommendation~(KGCR) model, in which we implement the deconfounded user preference learning to resolve the negative impacts of structural information and adopt counterfactual inference to eliminate the bias in the user-item similarity scoring. 
    \item Conducting extensive experiments on three datasets, we demonstrate the state-of-the-art performance of our proposed KGCR and its effectiveness in optimizing the user-item similarity scoring. 
\end{itemize}
\begin{table}
  \centering
  \setlength\tabcolsep{3.5pt}
  \renewcommand\arraystretch{1}
  \caption{Key notations and descriptions.}
  \begin{tabular}{l|l}
    \hline
    Notation&Description\\
    \hline
    \hline
    $\mathcal{G}_{kg}$ & knowledge graph (KG). \\
    $O$, $M$, and $N$ & number of users, items, and attributes, respectively. \\
    $\mathbf{u}\in\mathbb{R}^{1 \times D}$ & representation of user preference on attribute\\
    $\mathbf{i}\in\mathbb{R}^{1 \times D}$ & representation of item characterized by attributes. \\
    $\mathbf{u}_{cf}\in\mathbb{R}^{1 \times D}$ & user id embedding.\\
    $\mathbf{i}_{cf}\in\mathbb{R}^{1 \times D}$ & item id embedding. \\
    $\tilde{\mathbf{k}}\in [0,1]^{1 \times M}$ & distribution of items interacted by user. \\
    $\tilde{\mathbf{a}}_i\in [0,1]^{1 \times N}$ & distribution of the attributes characterizing item $i$.\\
    $\mathbf{e}_{(\cdot)} \in\mathbb{R}^{1 \times D}$ & embeddings of nodes and edges in the KG. \\
    $l$ & the number of graph convolutional layer.\\
    $S_{u,i,a}$ & similarity score between user and item.\\
    $S_{i,a}$ &  affinity between item and user interacted attributes.\\
    $S_{u,i}^{cf}$ & similarity score between user and item embeddings.\\
    $\mathcal{O}^+$, $\mathcal{O}^-$ & set of the positive and negative user-item pairs.\\
    $y_{ui}$ & prediction of interaction between user $u$ and item $i$.\\
    \hline
  \end{tabular}
    \label{table_annotation}
\end{table}
\section{Problem Formulation}
Given the user-item interactions in the history, CF-based recommendation model is designed to represent the users and items with the latent vectors, and predict their interactions according to the similarity between their representations~\cite{MMGCN,HS-GCN}. Formally, let $\mathcal{U}$ and $\mathcal{I}$ separately be the sets of $O$ users and $M$ items. The CF-based recommendation model can be formulated as:
\begin{equation}
    S^{cf}_{ui} = f_{cf}(\mathbf{u}_{cf},\mathbf{i}_{cf}),
\end{equation}
where $\mathbf{u}_{cf}$ and $\mathbf{i}_{cf}$ are the representations of user $u\in \mathcal{U}$ and item $i\in \mathcal{I}$, respectively. And, $S^{cf}_{ui}$ is the predicted score of interaction between $u$ and $i$, measuring how likely item $i$ will be interacted by user $u$ based on CF signal. In addition, the score function $f_{cf}(\cdot)$ can be implemented with the inner product~\cite{MF} or neural networks~\cite{NCF}. 

KG, as a side-information, can be introduced into the CF-based models~\cite{KGR1,KGR2,KGR3} to supplement the interaction information. It is a heterogeneous graph consisted by the triples~(\textit{e.g.,} \textit{Titanic}-\textit{directed\_by}-\textit{James Cameron}). Formally, KG can be represented as $\mathcal{G}_{kg}=\{(h, r, t)| h,t\in\mathcal{E}, r\in\mathcal{R}\}$, where $\mathcal{E}$ and $\mathcal{R}$ are the sets of entities and relations in the KG, respectively. Each triplet describes that there is a relationship $r$ from head entity $h$ to tail entity $t$. 
By mapping the items into the entities in the KG, the external knowledge learned from the KG could be injected into the representations of users and items. We list the important symbols in Table~\ref{table_annotation} to clarify their descriptions.

In this work, we explore to learn the user and item representations based on the attributes and estimate the user-item similarity $S_{u,i,a}$ at the fine-grained level, formally, 
\begin{equation}
    S_{u,i,a} = f(\mathbf{u}, \mathbf{i}),
    \text{where}\ \mathbf{u} = U(\mathcal{O}^+_u, \mathcal{G}_{kg}).
\end{equation}
Wherein, $f()$ and $U()$ are the similarity scoring and user preference modeling functions, respectively. And, the $\mathcal{O}^+_u$ is the historical interactions of user $u$ and $\mathcal{G}_{kg}$ denotes the knowledge graph.  
The task can be formulated as:
\begin{itemize}
    \item \textbf{Input}: the set of user-item interactions $\mathcal{O}^+=\{(u,i)|u\in\mathcal{U}, i\in\mathcal{I}\}$ and knowledge graph $\mathcal{G}_{kg}$.
    \item \textbf{Output}: the user-item interaction prediction $y_{ui}$ by integrating the user-item CF score $S^{cf}_{ui}$ and attribute-based similarity score $S_{u,i,a}$.
\end{itemize}

\section{Methodology}

\subsection{Causal Graph Construction and Analysis}
Causal graph is a directed acyclic graph, which offers us a tool to illustrate the causal-effect relation between the variables~\cite{Why}. In the causal graph, the nodes are used to represent the variables and the edges reflect the causal relation between them. 
In the following, we first introduce the causal graph constructed for our task. Then, we describe its nodes~(variables) and edges~(relations) in the causal-effect language and present the motivations of our designs.  

As shown in Figure~\ref{fig:figure2_1}, we construct the causal graph consisting of five variables: $U, I, A, K, S$. 
Noted that we omit the CF signal in this causal graph to emphasize the generation of user preference on attribute and the attribute-based similarity score. 
In particular, 
\begin{itemize}
    \item $U$ represents the user preference on attribute. For each user, its specific value $\mathbf{u}$ can be used to measure her/his interests to the item in a fine-grained level. 
    \item $I$ denotes the representation of item characterized by the attributes. For the item, the specific value $\mathbf{i}$ is learned from its related entities and relations in the KG. 
    \item $A$ represents the user's interacted attributes. 
    Here, we slightly abuse the notation $\mathbf{a}=\{\mathbf{a}_1, \mathbf{a}_{2}, \dots, \mathbf{a}_{A'}\}$ to denote the collection of the attributes interacted by the specific user. Whereinto, $\mathbf{a}_{j}$ is the representation of $j$-th attribute and $A'$ is the number of interacted attributes. 
    \item $K$ describes the structure information of items observed by user. Its specific value $\mathbf{k}=\{\mathbf{k}_1,\dots,\mathbf{k}_{K'} \}$ is a collection of $K'$ item' structure information. Thereinto, $\mathbf{k}_{j}\in[0,1]^{1 \times N}$ is $j$-th item's distribution over all $N$ attributes, reflecting how the attributes are entangled by $j$-th item. 
    \item $S$ whose specific value $s\in\mathbb{R}$ is the score to measure the attribute-based similarity of user-item pair.
\end{itemize}
With these variables, we next describe the directed edges in the causal graph, along which the successor nodes are affected by the ancestor nodes. In particular,
\begin{itemize}
    \item $K \rightarrow A:$ the structure information decides the exposure of attributes for the user. Since there is no observed record of user-attribute pairs, the collection of user's interacted attributes relies on the structure of items they consumed before. 
    \item $(K, A) \rightarrow U:$ the user preference on attribute is caused by the representations of attributes and structure information. Since the attribute cannot be exposed to users independently, we suggest the user preference learning is affected not only by \textit{which attributes are exposed to users} but by \textit{how they are entangled by the items which the user observed}. 
    \item $(A, U, I) \rightarrow S:$ the edges illustrate that the similarity score $S$ is affected by $U$, $I$, and $A$. As for $U$ and $I$, their similarity is the desired score to measure the user's interests to the item. While, the interacted attributes~$A$ may connect with the candidate items in the KG, causing the bias score via the shortcuts.
\end{itemize}
Applying the causal theory to the constructed causal graph, we observe that the structure information~($K$) as a confounder brings the spurious relation between the user preference~($U$) and user interacted attributes~($A$), which probably causes the suboptimal user representation. Furthermore, the edge $A \rightarrow S$ accounts for the direct effects on the score, which probably results in the bias issue for interaction prediction. 
\subsection{Knowledge Graph based Causal Recommendation}
According to our proposed causal graph, we conclude that the reasons causing the issues mainly come from two fork structures~(\textit{i.e.,} $A \leftarrow K \rightarrow U$ and $U \leftarrow A \rightarrow S$), which hinder the user preference learning and bias the similarity scores, respectively. Therefore, we do the intervention on $A$ for deconfounded user preference modeling and adopt the counterfactual inference to debiased the similarity scoring. 

\subsubsection{\textbf{Causal Intervention}}
As the before mentioned, $K$ acts as the \textit{confounder} in our proposed causal graph, opening a backdoor path: $A \leftarrow K \rightarrow U$, which both affects (or causes) either $A$ or $U$. It leads to spurious correlations by only learning from the likelihood $P(U|A)$. 
To resolve its negative impact on the user preference learning, we use \textit{do-calculus} operator to conduct the intervention on $A$. As shown in Figure~\ref{fig:figure2_2}, the intervention could be visually treated as cutting off on the edge pointing to $A$ in the causal graph. By interrupting the edge $K \rightarrow A$, we can remove the effect of $K$ to $A$, and then capture the user preference on attribute. Accordingly, we perform the backdoor adjustments to “virtually” achieve the formulation:
\begin{equation}
\begin{split}
    S_{u,i,a}&=
    P(S|U=\mathbf{u},I=\mathbf{i}, do(A=\mathbf{a}))\\
    &=\sum_{k\in\mathcal{K}}P(S|U(\mathbf{a}, \mathbf{k}),\mathbf{i},\mathbf{a})P(\mathbf{k})
    \label{equ:equ_3}
\end{split}
\end{equation}
where $P(\mathbf{k}) = \{P(\mathbf{k}_1), \dots, P(\mathbf{k}_{K'})\}$ reflects how likely the item associated its structure information observed by user. $U(\cdot)$ outputs the representation of user preferences based on their interacted attributes and structure information of items. 

Within Eq.~\ref{equ:equ_3}, we force the user interacted attributes fairly interact with all items to implement the intervention. As such, it is capable of opening their entanglement caused by the items. 
Thereafter, we can estimate Eq.~\ref{equ:equ_3} with:
\begin{equation}
    \begin{split}
        &P(S|U=\mathbf{u},I=\mathbf{i}, do(A=\mathbf{a})) \\
        = &\sum_{k\in\mathcal{K}}P(S|U(\mathbf{a}, \mathbf{k}),\mathbf{i},\mathbf{a})P(\mathbf{k})\\
        = &\sum_{k\in\mathcal{K}}f(U(\mathbf{a}, \mathbf{k}),\mathbf{i},\mathbf{a})P(\mathbf{k}),
    \end{split}
\end{equation}
where $f(\cdot)$ is the function to score the similarity between the user and item. The detail of its implementation will be described in the next section. 

Intuitively, we can estimate the similarity score by calculating the expected value of function $f(\cdot)$ of $\mathbf{k}$. However, the size of sample space $\mathcal{K}$ could be infinite in practice and expensive to conduct the computation on each possible value in $\mathcal{K}$. 
To overcome this drawback, we follow the prior work~\cite{wenjie,yangxun} and adopt the Normalized Weighted Geometric Mean~(NWGM)~\cite{NWGM} to approximate the formulation, as 
\begin{equation}
    P(S|U=\mathbf{u},I=\mathbf{i}, do(A=\mathbf{a}))
    \approx f(U(\mathbf{a}, (\sum_{k\in\mathcal{K}}\mathbf{k}P(\mathbf{k}))),\mathbf{i},\mathbf{a})
    \label{equ:eq5}
\end{equation}
It can approximately take outer sum $\sum_{k\in\mathcal{K}}$ into the function $U(\cdot)$ designed for the representation of user preference, so as to alleviate the computational burden. 
It is worth noting that if $f(\cdot)$ is a linear function with a random variable $X$ as the input, then $\mathbb{E}[f(X)] = f(\mathbb{E}[X])$ holds under any probability distribution $P(X)$. The proof can be found in the prior studies~\cite{proof1,proof2}. 
Thus, $U(\cdot)$ and $f(\cdot)$ should be implemented with the linear function to minimize the error of approximation. 
Beyond this, we deep into the $U(\cdot)$ and explore its inputs $\mathbf{k}$. 
According to the description of variable $K$, the specific value $\mathbf{k}$ consists of $K'$ vectors~(\textit{i.e.,} $\{\mathbf{k}_1,\dots,\mathbf{k}_{K'}\}$), each of which describes the distribution of one item over the attributes. 
Therefore, we can decompose $\mathbf{k}$ into two components. The first component is the distribution over all $M$ items, denoted as $\tilde{\mathbf{k}}\in [0,1]^{1 \times M}$, reflecting which items are sampled. And, the second component is a matrix $\mathbf{K}\in\mathbb{R}^{M \times N}$, where each row describes the corresponding item's distribution over attributions. 
As shown in Figure~\ref{fig:figure1}, given the item collection [\textit{Clockstoppers}, \textit{Aliens II}, \textit{Titanic}] and attribute collection [\textit{Micheal Biehn}, \textit{Fiction}, \textit{James Cameron}], $\mathbf{k}$ can be represented as the distribution $\tilde{\mathbf{k}}=[0,1,0]$ over items and matrix $\mathbf{K}=[[1,1,0],[1,1,1], [0,0,1]]$ over the attributes.
Accordingly, Eq.~\ref{equ:eq5} can be rewritten as,
\begin{equation}
    \begin{split}
    &P(S|U=\mathbf{u},I=\mathbf{i}, do(A=\mathbf{a}))\\
    \approx& f(U(\mathbf{a}, (\sum_{k\in\mathcal{K}}\mathbf{k}P(\mathbf{k}))),\mathbf{i},\mathbf{a})\\
    = & f(U(\mathbf{a}, \sum_{\tilde{\mathbf{k}}\in\tilde{\mathcal{K}}}(\tilde{\mathbf{k}}P(\tilde{\mathbf{k}})), \mathbf{K}), \mathbf{i},\mathbf{a}).\\
    \end{split}
\end{equation}
As such, we can change the distribution $\tilde{\mathbf{k}}$ in the counterfactual world to enforce $\mathbf{a}$ fairly exposed to the different structure information of item, instead of perturbing the connection between items and attributes in the KG. 
Assuming that each item has equal opportunity $\frac{1}{M}$ to be the interacted by users, we derive the formulation:
\begin{equation}
    \begin{split}
    &P(S|U=\mathbf{u},I=\mathbf{i}, do(A=\mathbf{a}))\\
    \approx& f(U(\mathbf{a}, \mathbb{E}(\tilde{\mathbf{k}}), \mathbf{K}), \mathbf{i},\mathbf{a}),\\
    =& \frac{1}{M}f(U(\mathbf{a}, \mathbf{K}), \mathbf{i},\mathbf{a}),\\
    =& \frac{1}{M}f(U(\tilde{\mathbf{a}}, \mathbf{A}, \mathbf{K}), \mathbf{i},\mathbf{a}).\\
    \end{split}
    \label{equ:equ7}
\end{equation}
Whereinto, we also decompose $\mathbf{a}$ into $\tilde{\mathbf{a}}\in [0,1]^{1 \times N}$ and $\mathbf{A}\in\mathbb{R}^{N \times D}$, which represent the distribution of interacted attributes and the matrix of all attributes' representations, respectively. 
\begin{figure}
	\centering
    \includegraphics[width=0.4\textwidth]{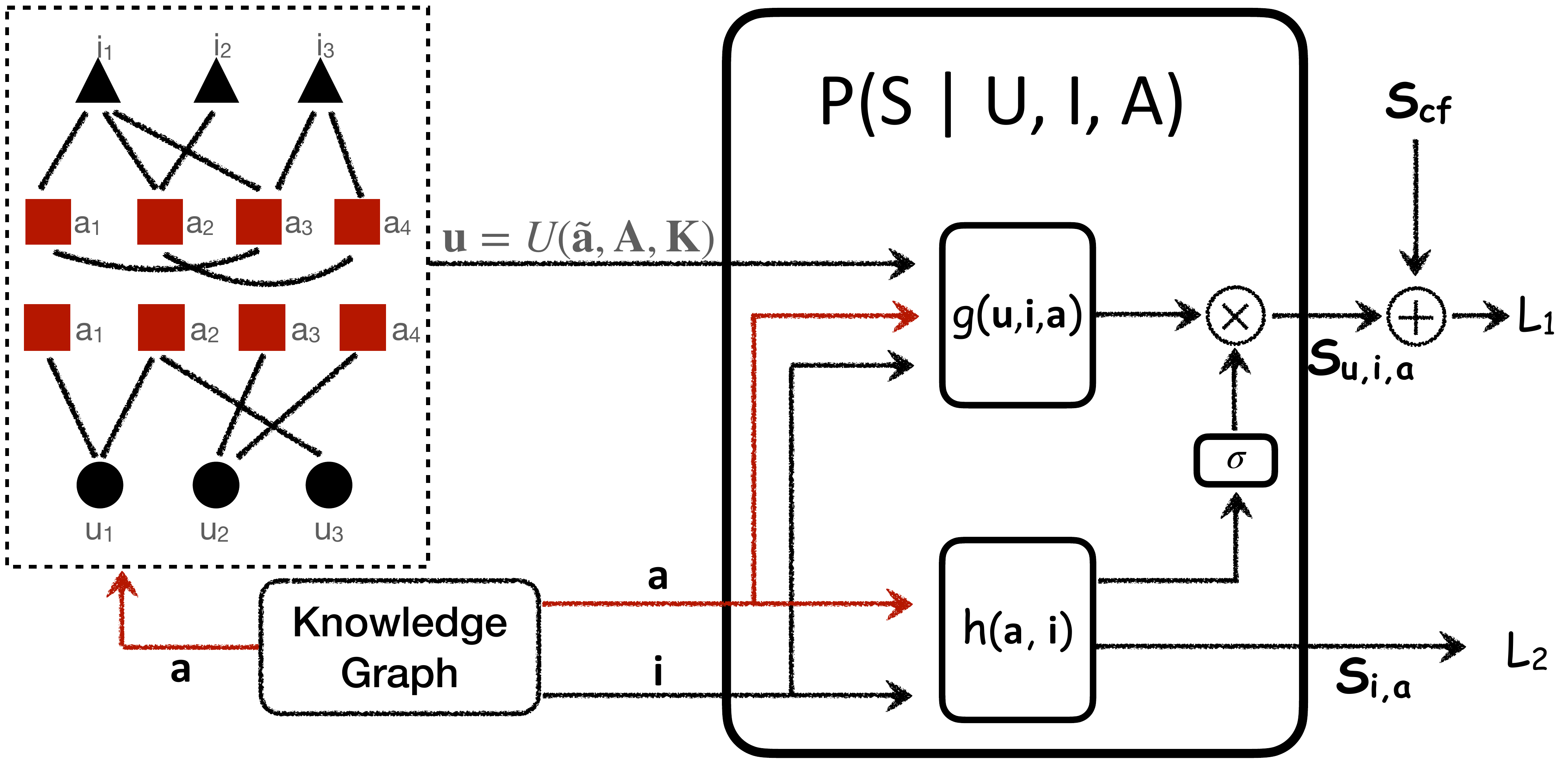}
    \caption{Illustration of our proposed model.}
	\label{fig:figure3}
\vspace{-5pt}
\end{figure}
\subsubsection{\textbf{Model Implementation}}
In this part, we present our proposed model KGCR to implement deconfounded user preference learning~$U(\cdot)$ and the similarity scoring~$f(\cdot)$, as shown in Figure~\ref{fig:figure3}. 
In particular, we first extract the knowledge information from the KG to initialize the item and attribute representations.
We perform the TransE model~\cite{TransE} on the KG due to its efficiency and robustness. More specifically, it learns the embeddings of entities and relations by minimizing the following loss function,
\begin{equation}
    \mathcal{L}_{KG} = \sum_{(h,r,t)\in\mathcal{T}}\sum_{(h',r,t')\in\mathcal{T}'}[\gamma + d(\mathbf{e}_h+\mathbf{e}_r, \mathbf{e}_t) - d(\mathbf{e}_{h'}+\mathbf{e}_r, \mathbf{e}_{t'})]_{+},
\end{equation}
where $[x]_{+}$ denotes the positive part of $x$. $(h, r, t)$  and $(h', r, t')$ represent the samples from the sets of triplets $\mathcal{T}$ and $\mathcal{T}'$, respectively. Different with $\mathcal{T}$ collected from the KG, $\mathcal{T}'$ consists of the broken triplets constructed by randomly replacing one entity in a real-world triplet. In addition, $\gamma$ is the margin hyperparameter. And $d()$ is implemented by Euclidean distance function to estimate the distance of the input representation pairs in a latent space.
\\\noindent\textbf{Implementing} $U(\cdot)$. 
Observing the derived function $U(\cdot)$ in Eq~\ref{equ:equ7}, we find that the user representation is depended on the representation of attributes~($\mathbf{A}$), distribution of interacted attribution~($\tilde{\mathbf{a}}$), and the structure information of KG~($\mathbf{K}$). 
To learn the representation of the user preference and attribute, we propose to construct two bipartite graphs: the item-attribute and user-attribute graphs, and separately conduct the graph convolutional operations on two bipartite~\cite{IMPGCN}. 
Considering that $U(\cdot)$ should be a linear function, we implement the graph convolutional operations, as,
\begin{equation}
    \mathbf{p}^{(l+1)} = \sum_{q\in\mathcal{N}_p}{\frac{1}{\sqrt{|\mathcal{N}_p||\mathcal{N}_q|}}\mathbf{q}^{(l)}},
\end{equation}
where $\mathbf{p}^{(l+1)}$ is the representation of $p$-th node at $(l+1)$-th graph convolutional layer. $|\mathcal{N}_p|$ and $|\mathcal{N}_q|$ represent the number of neighbors of nodes $p$ and $q$, respectively. 
By performing the operations, we aggregate the structure information into the attribute representation, and then model the fine-grained user preference based on the interacted attributes associated with the structure information. 
\\\noindent\textbf{Implementing} $f(\cdot)$. 
Recalling our proposed causal graph, we suggest that the similarity score comes from the user preference to item~(\textit{i.e.,} $(U, I) \rightarrow S$) and affinity between the item and user's interacted attributes~(\textit{i.e.,} $(A, I) \rightarrow S$). Hence, we define the score function $f(\cdot)$ as:
\begin{equation}
    S_{u,i,a} = f(\mathbf{u}, \mathbf{i}, \mathbf{a}) = g(\mathbf{u}, \mathbf{i})\sigma(h(\mathbf{i}, \mathbf{a})),
\end{equation}
where $\mathbf{u}\in\mathbf{U}$ is the user representations from function $U(\cdot)$ and $\sigma(\cdot)$ denotes the sigmoid function. In addition, $g(\cdot)$ and $h(\cdot)$ are the functions to separately model $(U, I) \rightarrow S$ and $(A, I) \rightarrow S$. 
This design, named Multiplication fusion strategy, is inspired by the prior studies~\cite{cadene2019rubi, wenjie}, which provides non-linearity for sufficient representation capacity of the fusion strategy. 
For simplicity, we conduct the inner product on the input vectors to implement $g(\cdot)$, formally,
\begin{equation}
    g(\mathbf{u}, \mathbf{i}) = \mathbf{u} \cdot \mathbf{i}^\intercal,
\end{equation}
where $\mathbf{i}\in\mathbf{I}$ is the item representation. As for the function $h(\cdot)$, we formulate it as,
\begin{equation}
    S_{i, a} = h(\mathbf{i}, {\mathbf{a}}) = \mathbf{i} \cdot \overline{\mathbf{a}}^\intercal, 
    \text{where}\ \overline{\mathbf{a}} = \frac{1}{M}\sum_{a\in A}\mathbf{a},
\end{equation}
where we empirically represent the interacted attributes $ \overline{\mathbf{a}}$ with the mean of their representations. 

In addition to the similarity score based on the knowledge information, we learn the id embeddings of users and items and score the user-item similarity based on the collaborative information with:
\begin{equation}
    S_{u,i}^{cf} = \mathbf{u}_{cf} \cdot \mathbf{i}_{cf}^\intercal,
\end{equation}
where we use $\mathbf{u}_{cf}\in\mathbf{U}_{cf}$ and $\mathbf{i}_{cf}\in\mathbf{I}_{cf}$ to denote the user and item id embeddings based on the collaborative information.
\subsection{Model Optimization}
To optimize the proposed model, we adopt Bayesian Personalized Ranking~(BPR) loss~\cite{BPR} in this work. In particular, if an item $i$ has been interacted by user $u$, then it is assumed that $u$ prefers $i$ over her/his unobserved item $j$. Incorporating the attribute-based and collaborative-based similarity scores~\cite{GRCN,HUIGN}, we formulate BPR loss as:
\begin{equation}
    \mathcal{L}_1 = \sum_{(u, i, j)\in\mathcal{O}}{-\ln\sigma(S_{u,i,a} + S^{cf}_{u,i} - S_{u,j,a} - S^{cf}_{u,j})},
    \label{Eq14}
\end{equation}
where $\mathcal{O}=\{(u,i,j)|(u,i)\in\mathcal{O}^+, (u,j)\in\mathcal{O}^-\}$ denotes the training set consisted of the positive~(observed) and negative~(unobserved) user-item pairs. 

Moreover, we also devise a loss function to supervise the output of $h(\cdot)$, as
\begin{equation}
    \mathcal{L}_2 = \sum_{(u, i, j)\in\mathcal{O}}max(0, \sigma(S_{i,a})-\sigma(S_{j,a})-m),
\end{equation}
where $m$ is a pre-defined margin value to control the difference between the predicted values of the positive and negative pairs. This is based on the assumption that the representation of positive item should be more similar with that of user's interacted attributes. Empirically, we find that the difference should be limited to a pre-defined range, in order to achieve the promising results. 

Combining these two loss functions, we obtain the objective function:
\begin{equation}
    \mathcal{L} = \mathcal{L}_1 + \alpha\mathcal{L}_2 + ||\Theta||_2,
\end{equation}
where $\alpha$ is a hyper-parameter to balance the two loss functions and $||\Theta||_2$ is the regularization term. 

\subsection{Counterfactual Inference Strategy}
\begin{figure}
	\centering
    \subfigure[Counterfactual World]{
    \includegraphics[width=0.2\textwidth]{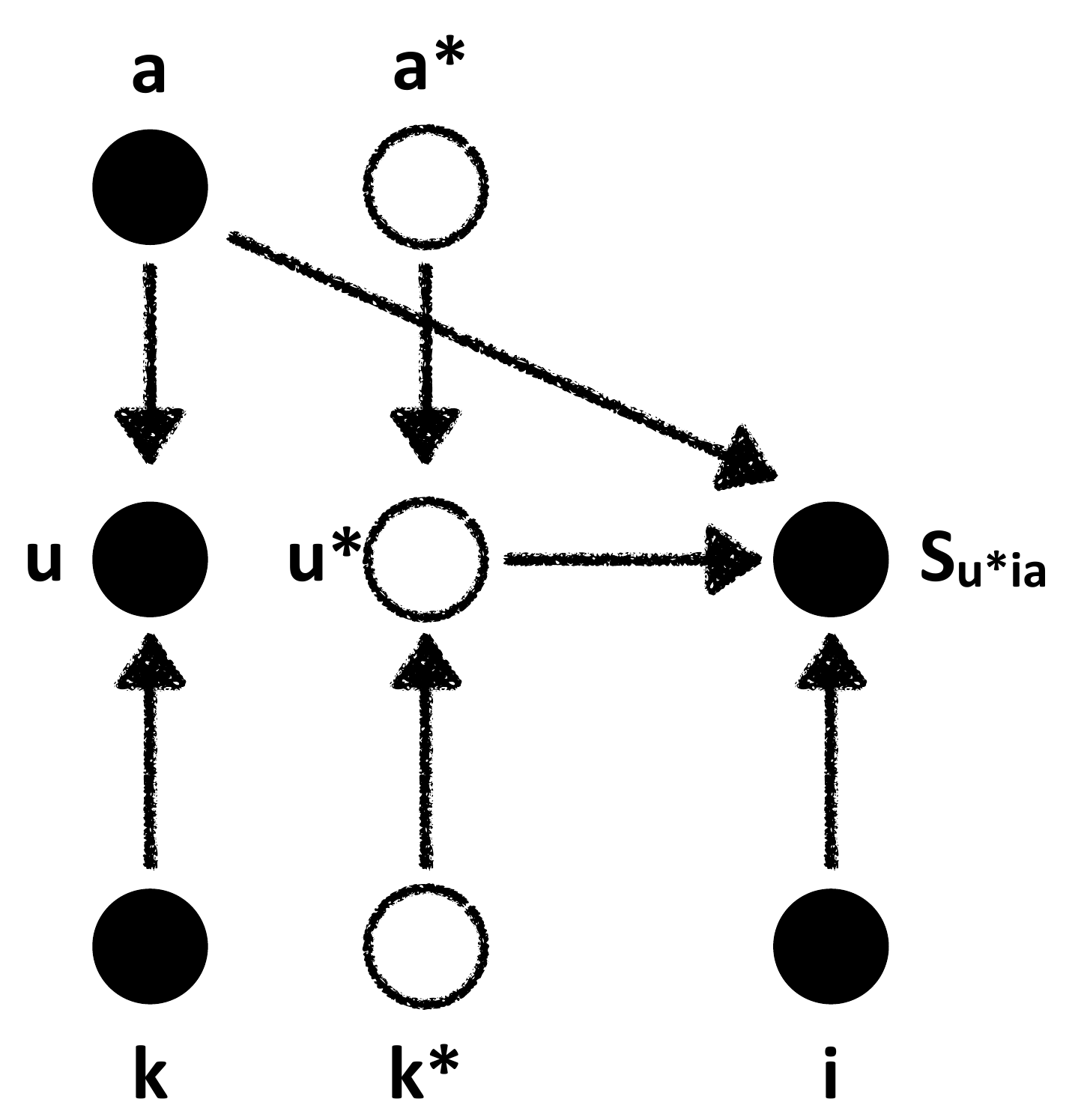}
    \label{fig:figure4_a}
    }
    \subfigure[Reference Situation]{
    \includegraphics[width=0.2\textwidth]{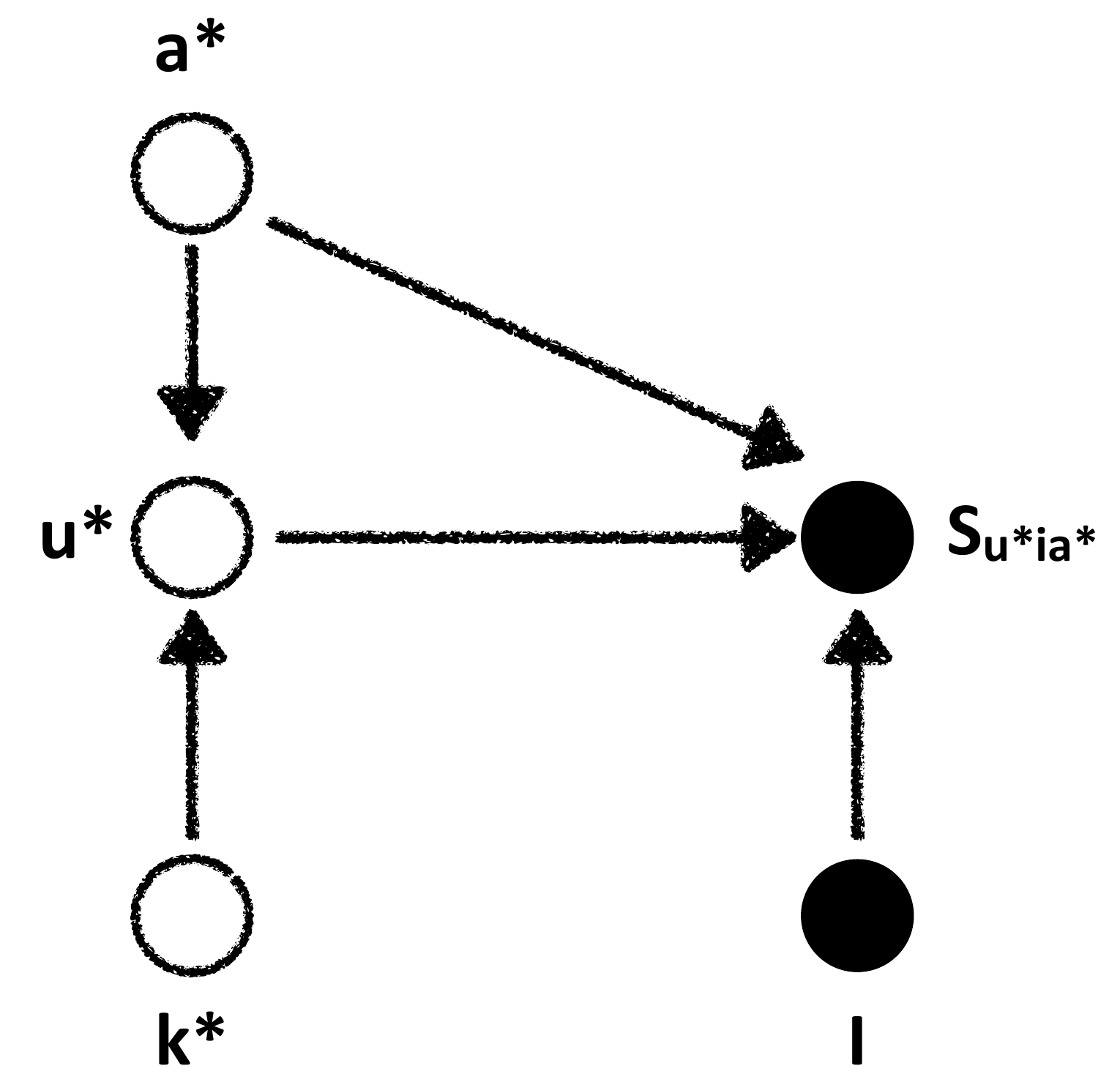}
    \label{fig:figure4_b}
    }
    \caption{Illustration of Counterfactual Inference.}
\end{figure}
Although we eliminate the impact of the confounder $A$ in the user preference learning, we still suffer from the bias issue in the user-item similarity score as the before mentioned. To resolve this problem, we opt to do the counterfactual inference~\cite{DIE} by answering the question: \textit{what the similarity score would be, if we only consider the attribute-based representations of users and items}. 

\begin{algorithm}[t]
\caption{Knowledge Graph-based Causal Recommendation}
\algorithmicrequire{ $\mathcal{O}^+=\{(u,i)|u\in\mathcal{U}, i\in\mathcal{I}\}$, $\mathcal{G}_{kg}=\{(h, r, t)| h,t\in\mathcal{E}, r\in\mathcal{R}\}$}\\
\algorithmicensure{ $\Theta=\{\mathbf{U}, \mathbf{I}, \mathbf{U}_{cf}, \mathbf{I}_{cf}\}, \{S^{cf}_{u,i}, S_{u,i,a}\}$}\\
\begin{algorithmic}[1]
\STATE Initialization all model parameters in $\Theta$;
\STATE Calculate embeddings $\mathbf{E}$ of KG \textit{w.r.t.} Eq.8;
\STATE Initialization $\mathbf{I}$ and $\mathbf{A}$ with $\mathbf{E}$;
\WHILE {stopping criteria is not met} 
    \FOR{each layer $l\in\{1, 2, \dots, L\}$}
        \STATE obtain $\mathbf{u}^l$ and $\mathbf{a}^l$ \textit{w.r.t.} Eq.9;
    \ENDFOR
    \FOR{each layer $l\in\{1, 2, \dots, L\}$}
        \STATE obtain $\mathbf{i}^l$ and $\mathbf{a}^l$ \textit{w.r.t.} Eq.9;
    \ENDFOR
    \STATE $g(\mathbf{u}, \mathbf{i}) = \mathbf{u} \cdot \mathbf{i}^\intercal$;
    \STATE $S_{i, a} = h(\mathbf{i}, {\mathbf{a}}) = \mathbf{i} \cdot \overline{\mathbf{a}}^\intercal, 
    where\ \overline{\mathbf{a}} = \frac{1}{M}\sum_{a\in A}\mathbf{a}$;
    \STATE $S_{u,i,a} = f(\mathbf{u}, \mathbf{i}, \mathbf{a}) = g(\mathbf{u}, \mathbf{i})\sigma(h(\mathbf{i}, \mathbf{a}))$;
    \STATE $S_{u,i}^{cf} = \mathbf{u}_{cf} \cdot \mathbf{i}_{cf}^\intercal$;
    \STATE Calculate $\mathcal{L}$ \textit{w.r.t.} Eq. 16;
    \STATE Update parameters \textit{w.r.t.} $\Theta$ with Adam
\ENDWHILE
\RETURN $\Theta$, $S^{cf}_{u,i}$, $S_{u,i,a}$
\end{algorithmic}
\label{algorithm1}
\end{algorithm}

According to the prior work~\cite{BoW,DecRS}, the counterfactual inference is able to estimate what the descendant variables would be if the value of one treatment variable changed from its value in the factual world. Leveraging this counterfactual thought, we can estimate NDE of variable $A$ to $S$ in the counterfactual world and remove it from TE in the factual world to answer the above question. 
In particular, TE is the causal effect of the treatment variable~(\textit{e.g.,} $A$) on the response variable~(\textit{e.g.,} $S$), which can be estimated by the change of response variable when the treatment variable changed from the reference value to the expected value. 
Formally, TE can be estimated by the following formulation:
\begin{equation}
    TE = S_{u,i,a} - S_{u^*,i,a^*},
\end{equation}
where, $u^*$ and $a^*$ are the reference values of the variables $U$ and $A$. 

Different from TE, NDE is the change of the response variable when only changing the treatment variable on the direct path, as illustrated in Figure~\ref{fig:figure4_a}. 
In the counterfactual world, we formulate the NDE as,
\begin{equation}
    NDE = S_{u^*,i,a} - S_{u^*,i,a^*},
\end{equation}
where $(u^*, i, a)$ is the counterfactual instance generated by conducting the \textit{do-calculus} operator on variable $U$, \textit{i.e.,} $do(U=u^*)$.

Following, by removing NDE from TE, we can obtain the debiased effect of user and item to the score, which is termed as the \textit{total indirect effect}~(TIE). Formally, TIE could be formulated as,
\begin{equation}
    \begin{split}
    TIE &= TE - NDE \\
    &= S_{u,i,a}  - S_{u^*,i,a^*} - S_{u^*,i,a} + S_{u^*,i,a^*}\\
    &= g(\mathbf{u}, \mathbf{i})\sigma(h(\mathbf{i}, \mathbf{a})) - g(\mathbf{u}^*, \mathbf{i})\sigma(h(\mathbf{i}, \mathbf{a}))\\
    &= (g(\mathbf{u}, \mathbf{i}) - {s}_i)\sigma(h(\mathbf{i}, \mathbf{a})).
    \end{split}
    \label{Equ18}
\end{equation} 
Thereinto, since the reference value $\mathbf{u}^*$ is independent with the specific user, ${s}_i$ could be computed by the mean of the output of $g(\mathbf{u}, \mathbf{i})$.

Finally, we incorporate the debiased similarity score with the collaborative score to predict the interactions between the users and items, formally, 
\begin{equation}
    y_{ui} = S^{cf}_{ui} + S^{kg}_{ui} = \mathbf{u}_{cf}^\intercal \cdot \mathbf{i}_{cf} +   (g(\mathbf{u}, \mathbf{i}) - {s}_i)\sigma(h(\mathbf{i}, \mathbf{a})).
    \label{equ:equ15}
\end{equation} 

\begin{table}
  \centering
  \small
  \renewcommand\arraystretch{1}
  \caption{Summary of the datasets. (U-I means the user-item interaction.) }
  \label{table_1}
  \begin{tabular}{p{8mm}|c|c|c|c}
    \hline
    & & Amazon-book & LastFM & Yelp2018 \\
    \hline
    \hline
    \multirow{3}*{\minitab[c]{U-I}}&\#Users & 70,679 & 23,566 & 45,919\\
    &\#Items & 24,915 & 48,123 & 45,538\\
    &\#Inter. &  847,733 & 3,034,796 & 1,185,068\\
    \hline
    \hline
    \multirow{3}*{\minitab[c]{KG}}&\#Entities & 88,572 & 58,266 & 90,961\\
    &\#Relations & 39 & 9 & 42\\
    &\#Triplets &  2,557,746 & 464,567 & 1,853,704\\
    \hline
  \end{tabular}
\end{table}

\begin{table}
  \centering
  \setlength\tabcolsep{3.5pt}
  \renewcommand\arraystretch{1.1}
  \caption{Overall performance comparison.}
  \begin{tabular}{c|cc|cc|cc}
    \hline
    \multirow{2}{*}{Model}&\multicolumn{2}{c|}{Amazon-book}&\multicolumn{2}{c|}{LastFM}&\multicolumn{2}{c}{Yelp2018}\\
    \cline{2-7}
    &Recall&NDCG&Recall&NDCG&Recall&NDCG\\
    \hline
    \hline
    CKE & 0.1343 & 0.0698 & 0.0736 & 0.0630& 0.0651 & 0.0414 \\
    RippleNet & 0.1336 & 0.0691 & 0.0791 & 0.0684 & 0.0664 & 0.0428\\
    \hline
    KTUP & 0.1369 & 0.0680 & 0.0783 & 0.0681& 0.0640 & 0.0420\\
    MKR & 0.1286 & 0.0676& 0.0743 & 0.0642 & 0.0634 & 0.0409\\
    \hline
    KGNN-LS & 0.1362 & 0.0560 & 0.0880 & 0.0642 & 0.0637 & 0.0402\\
    CKAN & 0.1442 & 0.0698 & 0.0812 & 0.0660 & 0.0604 & 0.0377\\
    KGAT & 0.1489 & 0.0799 & 0.0870 & 0.0744 & 0.0712 & 0.0443\\
    KGIN & \underline{0.1687} & \underline{0.0915} & \underline{0.0978} & \underline{0.0848} & \underline{0.0736} & \underline{0.0482}\\
    \hline
    KGCR & \textbf{0.1751} & \textbf{0.0949} & \textbf{0.1036} & \textbf{0.0883} & \textbf{0.0785} & \textbf{0.0518}\\
    \hline    
    \hline    
    \textit{Improv\%} & 3.79\% & 3.72\% & 5.93\% & 4.13\% & 6.66\% & 7.47\%\\
    \hline
  \end{tabular}
    \label{table_2}
\end{table}

\section{EXPERIMENTS}
\subsection{Experimental Settings}
\subsubsection{\textbf{Dataset}}
In our work, we do the experiments on Amazon-book, LastFM, and Yelp2018 datasets, which are released by KGAT and widely used by KG-based recommendation models. These datasets involve three different applications in real scenarios, including online shopping, music, and business recommendations. According to the description in KGAT, each of them contains the user-item interactions and knowledge graph consisted of multiple triplets, \textit{i.e.,} $\langle object, relation, subject \rangle$. In summary, we detail the information of datasets associated with the knowledge graph in Table~\ref{table_1}.

For each dataset, we use the training set provided in KGAT~\cite{KGAT} to optimize the parameters of our proposed model and baselines. 
In the training phase, we take each user-item interaction in the set and randomly select one negative item that is unseen by the user to form the training triple.  

\subsubsection{\textbf{Baselines}}
We compare KGCR with several state-of-the-art KG-based recommendation models\footnote{For fairness, we discard KV-MN model~\cite{KVMN}, since it is designed for sequential recommendation task.}, including,
\begin{itemize}
    \item \textbf{CKE}~\cite{CKE} As one of the representative KG-based recommendation models, CKE conducts TransR method to learn the knowledge information from the KG and injects the information into the id embeddings to supplement the CF-based recommendation. 
    \item \textbf{RippleNet}~\cite{Ripplenet} RippleNet represents the user preferences as the set of her/his interacted items in the KG and extends the set by propagating the preference along the edge of the KG, so as to leverage the external knowledge in the recommendation.  
    \item \textbf{KTUP}~\cite{KTUP} Treating the user-item interaction as one type of relation between entities~(\textit{i.e.,} user and item entity), KTUP jointly optimizes the task of recommendation and knowledge graph completion, which is able to transfer the information learned from KG into the user-item interaction prediction.
    \item \textbf{MKR}~\cite{MKR} To assist the recommendation model with KGE, MKR conducts the multi-task training strategy to simultaneously optimize KGE and recommendation tasks, in which a cross\&compress unit is devised to integrate the learned knowledge and collaborative information.
    \item \textbf{KGNN-LS}~\cite{KGNN-LS}  To generate user-specific item representations, this model converts the KG into user-specific graphs, and considers user preference on KG relations and label smoothness in the information aggregation phase.
    \item \textbf{CKAN}~\cite{CKAN} CKAN explicitly encodes the collaborative signals by collaboration propagation and applies a knowledge-aware attention mechanism to incorporate the collaborative information with knowledge information. 
    \item \textbf{KGAT}~\cite{KGAT} KGAT applies the graph attentive network on a holistic graph combined by the KG and user-item graph, which is able to explicitly construct the high-order relation among the user, item, and attributes. It encodes the collaborative and knowledge information into the user and item in an end-to-end manner. 
    \item \textbf{KGIN}~\cite{KGIN} This is a state-of-the-art KG-based recommendation model. It models the user intents behind user-item interactions with the knowledge information learned from the KG, so as to improve the performance and interpretability of recommendation.
\end{itemize}

\subsubsection{\textbf{Evaluation Metrics}}
For each user, we take all items she/he does not interact with during the training phase, including the ground-truth and negative items, as the candidate samples. Using the trained model, we score the interactions of the user-item pairs and rank them in descending order. Then, we adopt Recall@K and Normalized Discounted Cumulative Gain~(NDCG@K for short) to evaluate the effectiveness of top-K recommendation. By default, we set K=20 and report the average values of the above metrics for all users during the testing phase. 

\subsubsection{\textbf{Parameter Settings}}
With the help of Pytorch\footnote{https://pytorch.org/.} and torch-geometric package\footnote{https://pytorch-geometric.readthedocs.io/.}, we implement the baselines and our proposed model. 
We utilize Xavier~\cite{Xavier} and Adam~\cite{Adam} algorithms in the experiments to initialize and optimize the parameters of the models. For fairness, we set the dimensions of the id embedding and entity representation as 64 for all models, and adopt the three-layer graph convolutional network to learn the id embeddings in KGCR and GCN-based models. 
In terms of the hyper-parameters, we tune the learning rate in range of $\{0.0001, 0.001, 0.01, 0.1\}$ and regularization weight in range of $\{0.0001, 0.001, 0.01, 0.1\}$. Besides, we employ the same early stopping strategy with KGAT, which stops the training if Recall@20 does not increase for 10 successive epochs. For the baselines, we do the same options and follow the designs (\textit{e.g.,} the number of graph convolutional layers and multilayer perceptron) in their articles to achieve the best performance.

\begin{table}
  \centering
  \setlength\tabcolsep{3.5pt}
  \renewcommand\arraystretch{1.1}
  \caption{Impact of Deconfounded User Preference and Debiased Similarity.}
  \begin{tabular}{c|cc|cc|cc}
    \hline
    \multirow{2}{*}{Model}&\multicolumn{2}{c|}{Amazon-book}&\multicolumn{2}{c|}{LastFM}&\multicolumn{2}{c}{Yelp2018}\\
    \cline{2-7}
    &Recall&NDCG&Recall&NDCG&Recall&NDCG\\
    \hline
    \hline
    KGCR & \textbf{0.1751} & \textbf{0.0949} & \textbf{0.1036} & \textbf{0.0883} & \textbf{0.0786} & \textbf{0.0518}\\
    \hline
    w/o DC & 0.1565 & 0.0847 & 0.0867 & 0.0737 & 0.0721 & 0.0469\\
    \hline
    w/o CI & 0.1716 & 0.0926 & 0.0996 & 0.0838 & 0.0774 & 0.0510\\
    \hline
  \end{tabular}
    \label{table_3}
\end{table}
\begin{table}
  \centering
  \setlength\tabcolsep{3.5pt}
  \renewcommand\arraystretch{1.1}
  \caption{Impact of Implementation of $U(\cdot)$}
  \begin{tabular}{c|cc|cc|cc}
    \hline
    \multirow{2}{*}{Model}&\multicolumn{2}{c|}{Amazon-book}&\multicolumn{2}{c|}{LastFM}&\multicolumn{2}{c}{Yelp2018}\\
    \cline{2-7}
    &Recall&NDCG&Recall&NDCG&Recall&NDCG\\
    \hline
    \hline
    KGCR & \textbf{0.1751} & \textbf{0.0949} & \textbf{0.1036} & \textbf{0.0883} & \textbf{0.0786} & \textbf{0.0518}\\
    \hline
    w/o U-A & 0.1703 & 0.0924 &0.0927 & 0.0742 & 0.0732 & 0.0476\\
    \hline
    w/o I-A & 0.1662 & 0.0901 & 0.0955 & 0.0747 & 0.0750 & 0.0487\\
    \hline
  \end{tabular}
    \label{table_4}
\end{table}

\subsection{Overall Performance Comparison}
To demonstrate the effectiveness of our proposed model, we start by doing the comparison between KGCR and the baselines \textit{w.r.t.} Recall@20 and NDCG@20. Specifically, we list the results on three datasets in Table~\ref{table_2}, where $Improv.\%$ represents the relative improvements of the best performing method (bolded) over the strongest baselines (underlined). 
Observing the table from the bottom to top, we have the following findings:
\begin{itemize}
    \item Without any doubts, our proposed model consistently achieves the best performance on the three datasets, which is able to justify the effectiveness of our proposed model. In particular, KGCR improves over the strongest baselines \textit{w.r.t.} Recall@20 by 3.79\%, 5.93\%, and 6.66\% in Amazon-Book, LastFM, and Yelp2018, respectively. We attribute the improvements to modeling the fine-grained user preference based on the knowledge information and supplement the collaborative information with the attribute-based similarity between the user and item. 
    \item Analyzing the results of our KGCR across the datasets, we find that its improvement on Yelp2018 is more significant than that on Amazon-book dataset. The reason might be from their different applications. Since the items from Yelp2018 involve more topics\footnote{https://www.yelp.com/dataset/documentation/main}, learning the finer-grained user preference on Yelp2018 could yield better performance than that on Amazon-book just constructed for the book recommendation, which is consistent with the findings in KGIN. 
    \item KGIN outperforms the other baselines by a margin over three datasets. We suggest that its advantages mainly come from its obtained user intents. Although KGIN hardly distinguishes the knowledge information from the collaborative one, the results of KGIN and KGCR verify the effectiveness of modeling fine-grained user intents/preferences in the recommendation method. 
    \item Comparing with the GCN-based models (\textit{i.e.,} KGNN-LS, CKAN, KGAT, KGIN, and KGCR), the other models are suboptimal in most cases. It indicates that the graph convolutional operation is capable of capturing the collaborative and knowledge information from the graph and then enhancing the representations of users and items. 
\end{itemize}

\subsection{Ablation Study}
To evaluate the designs in our proposed model, we first justify the effectiveness of the deconfounded user preference and debiased similarity score in KGCR. Then, we delve into them to test the implementation of functions $U(\cdot)$, $h(\cdot)$, and loss function $\mathcal{L}_2$.
\subsubsection{\textbf{Impact of Deconfounded User Preference and Debiased Similarity Score}} 
We devise two variants: $\text{KGCR}_{\text{w/o DC}}$ and $\text{KGCR}_{\text{w/o CI}}$, which remove the deconfounded user preference and counterfactual inference from KGCR, respectively. 
To implement $\text{KGCR}_{\text{w/o DC}}$, we ignore the user interacted attributes and estimate the user-item interaction on attributes. Thus, we replace the term $S_{u,i,a}$ in Equ.~\ref{Eq14} by:
\begin{equation}
    S_{u,i} = g(\mathbf{u}, \mathbf{i}) = \mathbf{u} \cdot \mathbf{i}^\intercal.
\end{equation}
As for $\text{KGCR}_{\text{w/o CI}}$, we omit the term $S_{u^*,i,a}$ in Equ.~\ref{Equ18} and obtain the interaction of user-item pair: 
\begin{equation}
    y_{ui} = \mathbf{u}_{cf} \cdot \mathbf{i}_{cf}^\intercal + g(\mathbf{u}, \mathbf{i})\sigma(h(\mathbf{i}, \mathbf{a})).
\end{equation}

\begin{table}
  \centering
  \setlength\tabcolsep{3.5pt}
  \renewcommand\arraystretch{1.1}
  \caption{Impact of Implementation of $h(\cdot)$.}
  \begin{tabular}{c|cc|cc|cc}
    \hline
    \multirow{2}{*}{Model}&\multicolumn{2}{c|}{Amazon-book}&\multicolumn{2}{c|}{LastFM}&\multicolumn{2}{c}{Yelp2018}\\
    \cline{2-7}
    &Recall&NDCG&Recall&NDCG&Recall&NDCG\\
    \hline
    \hline
    MLP & 0.1735 & 0.0933 & 0.1018 & 0.0876 & 0.0781 & 0.0515\\
    \hline
    Weighted & 0.1742 & 0.0942 & \textbf{0.1036} & \textbf{0.0885} & \textbf{0.0786} & \textbf{0.0518}\\
    \hline
    Mean & \textbf{0.1751} & \textbf{0.0949} & \textbf{0.1036} & 0.0883 & 0.0785 & \textbf{0.0518}\\
    \hline
  \end{tabular}
    \label{table_5}
\end{table}
\begin{table}
  \centering
  \setlength\tabcolsep{3.5pt}
  \renewcommand\arraystretch{1.1}
  \caption{Impact of Implementation of $\mathcal{L}_2$.}
  \begin{tabular}{c|cc|cc|cc}
    \hline
    \multirow{2}{*}{Model}&\multicolumn{2}{c|}{Amazon-book}&\multicolumn{2}{c|}{LastFM}&\multicolumn{2}{c}{Yelp2018}\\
    \cline{2-7}
    &Recall&NDCG&Recall&NDCG&Recall&NDCG\\
    \hline
    \hline
    LOG & 0.1630 & 0.0877 & 0.1008 & 0.0853 & 0.0772 & 0.0506\\
    \hline
    MAX & 0.1738 & 0.0924 & 0.1012 & 0.0858 & 0.0770 & 0.0509\\
    \hline
    KGCR & \textbf{0.1751} & \textbf{0.0949} & \textbf{0.1036} & \textbf{0.0883} & \textbf{0.0786} & \textbf{0.0518}\\
    \hline
  \end{tabular}
    \label{table_6}
\end{table}

Observing the results shown in Table~\ref{table_3}, we find that the results \textit{w.r.t.} NDCG and Recall are dramatically reduced after removing the two components on three datasets. It demonstrates the necessity of our designed deconfounded user preference learning and debiased similarity scoring in KGCR. 
Jointly analyzing Table~\ref{table_2}, $\text{KGCR}_{\text{w/o CI}}$ yields better results than KGIN. We suggest that the improvements come from the deconfounded user preference learning, while KGIN suffers from the negative influence of structure information to fine-grained user preference/intent modeling. It verifies the effectiveness of eliminating the confounder from user preference learning. Moreover, we observe that $\text{KGCR}_{\text{w/o DC}}$ outperforms the baselines in most cases. It justifies our argumentation that the bias issue challenges the KG-based recommendation. 
\subsubsection{\textbf{Impact of Implementation of $U(\cdot)$}}
After evaluation of deconfounded user preference, we further test the implementation of function $U(\cdot)$, which is at the core of deconfounded user preference learning. As mentioned before, we conduct the graph convolutional operations on the user-attribute~(U-A for short) and item-attribute~(I-A for short) graphs to represent the user preference on attribute. To justify our implementation, we separately discard U-A and I-A to devise the variants: $\text{KGCR}_{\text{w/o U-A}}$ and $\text{KGCR}_{\text{w/o I-A}}$\footnote{Discarding both of two graphs equals removing the user preference learning, that is $\text{KGCR}_{\text{w/o DC}}$ in Table 3.}. 

After removing the graph from the proposed model, we capture the representations of user nodes from a trainable matrix and optimize them by conducting the back-propagation operation under the supervision of user-item interactions, instead of the message propagation of GCN model. By comparing the variants with KGCR model, we have the following observations:


From their results listed in Table~\ref{table_4}, we observe that KGCR is superior to all variants on three datasets. It verifies the rationality and effectiveness of the implementation of function $U(\cdot)$. In addition, we find the performance is decreased when we remove I-A graph from function $U(\cdot)$. This indicates that the spurious relation caused by the item structure information hinders the user preference on attribute. More importantly, the better results of KGCR demonstrate that we eliminate the confounder by using the structure information extracted from I-A graph. Jointly analyzing Table~\ref{table_3}, $\text{KGCR}_{\text{w/o U-A}}$ unexpectedly outperforms $\text{KGCR}_{\text{w/o DC}}$, even U-A is removed. But, it is reasonable since we explicitly model the causal-effect relation $K \rightarrow A$, where $K$ constructs the statistic correlation between user preference and attributes via the back-door path $A \leftarrow K \rightarrow U$. We suggest that it quantitatively illustrates our constructed causal graph. 
\begin{figure}
	\centering
    \includegraphics[width=0.45\textwidth]{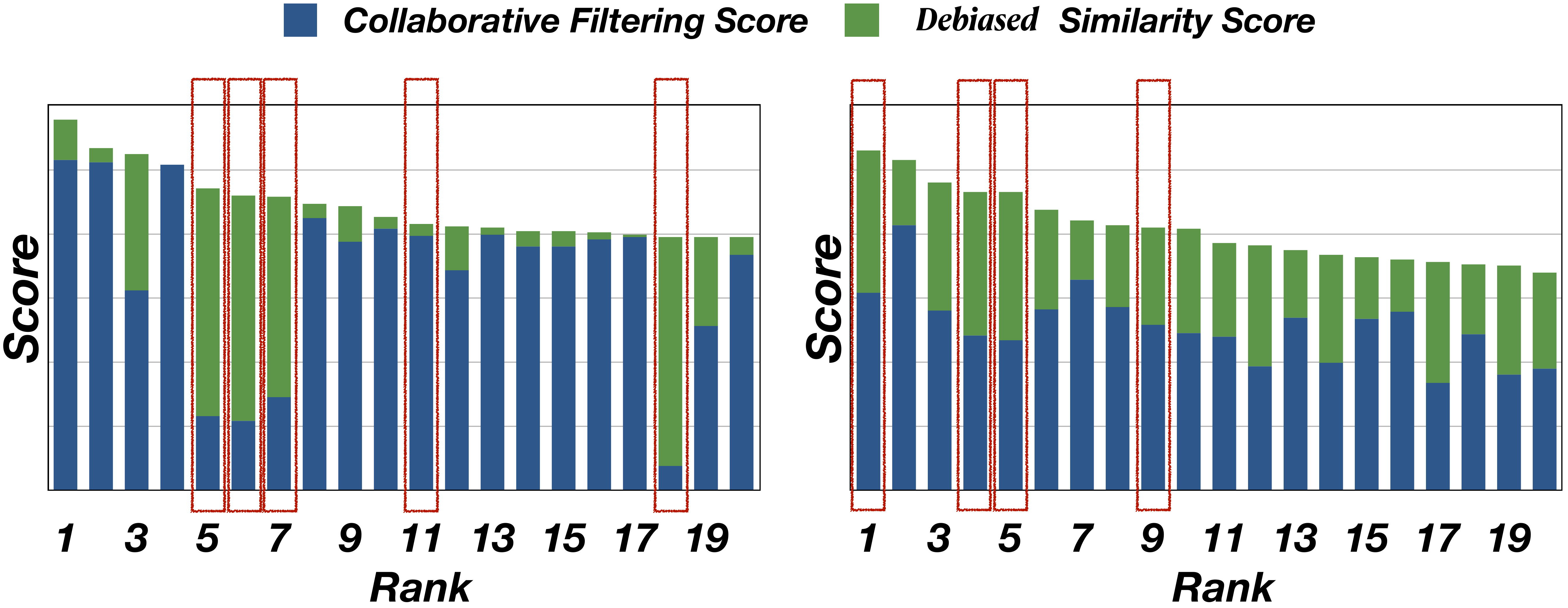}
    \caption{The predicted score of top 20 ranking of users \#47847 (left) and \#50428 (right). Best viewed in color.}
	\label{fig:figure5}
\end{figure}
\subsubsection{\textbf{Impact of Implementation of $h(\cdot)$}}
Exploring the function $h(\cdot)$, it is easy to find that the representation of user's interacted attributes plays a pivotal role in debiased similarity scoring, which directly affects the estimation of NDE of A to S. Thus, we adopt three different methods to model the representation and study their influence to the prediction. Beyond the mean value of the interacted attributes, we also utilize the multi-layer perceptron~(MLP) to optimize the representation and use the weighted average to adjust the different frequencies of interacted attributes. Conducting the experiments, we summarize their results in Table~\ref{table_5}, where MLP, Weighted, and Mean denote three methods, respectively. 

According to the results in Table~\ref{table_5}, Mean method achieves the comparable results to Weighted method in LastFM and Yelp2018 datasets, while slightly performing better than Weighted in Amazon-book dataset. The reason might be that the dense attributes per item on Amazon-book brings much more uninterested attributes to users, whereas the users tend to prefer several attributes of items in practice. Moreover, we observe MLP method performs poor across three datasets. We attribute the suboptimal performance to the overfitting problem. It is consistent with the findings in LightGCN~\cite{LightGCN} that performing multiple layers of nonlinear feature transformation may bring no benefits, even negative influence, to recommendation models.
\subsubsection{\textbf{Impact of implementation of loss function}}
Here, we explore the impact of loss function $\mathcal{L}_2$, which is used to supervise the similarity between the representations of the user's interacted attributes and item. We compare our KGCR with two different variants, formally, 
\begin{equation}
    \left\{
    \begin{split}
    &KGCR_{LOG}: \mathcal{L}_2 = log(\sigma(S_{i,a}-S_{j,a}))\\
    &KGCR_{MAX}: \mathcal{L}_2 = max(0, S_{i,a}-S_{j,a}-m)\\
    &KGCR: \mathcal{L}_2 = max(0, \sigma(S_{i,a})-\sigma(S_{j,a})-m)
    \end{split}
    \right.
\end{equation}
We summarize their results on three datasets in Table~\ref{table_6}. 

As expected, KGCR equipped with original $\mathcal{L}_2$ achieves the better results on three datasets. It can be attributed to that the function controls the difference between the scores of positive and negative pairs~(\textit{i.e.,} $S_{i,a}$ and $S_{j,a}$) in an appropriate range. Specifically, $\text{KGCR}_{\text{LOG}}$ underperforms the other two methods on all datasets. It might be that the positive pair is scored higher than the negative pair by a large margin. Comparing it with $\text{KGCR}_{\text{MAX}}$, we observe that, in most cases, the later one yields the better performance by limiting the margin between the scores. It justifies that the score of the positive pair is just slightly higher than that of the negative one in the real scenario. However,  $\text{KGCR}_{\text{MAX}}$ may result in the overlarge scores of all user-item pairs, which makes the difference vanish after conducting the sigmoid function.  

\begin{figure}
	\centering
	\subfigure[Recall on Amazon-book]{
    \includegraphics[width=0.22\textwidth]{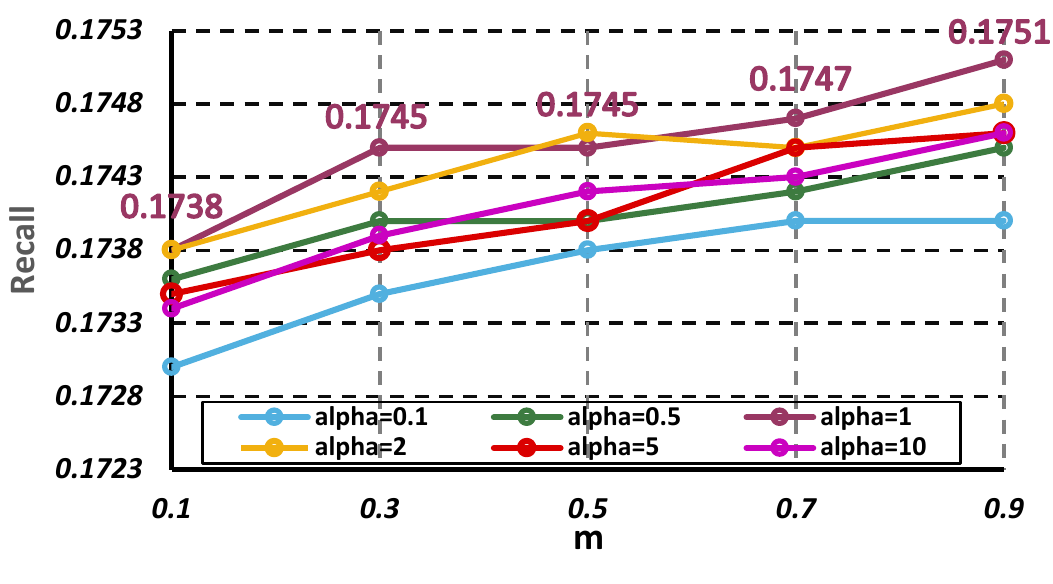}
	\label{fig:figure5_1}
    }
    \subfigure[NDCG on Amazon-book]{
    \includegraphics[width=0.22\textwidth]{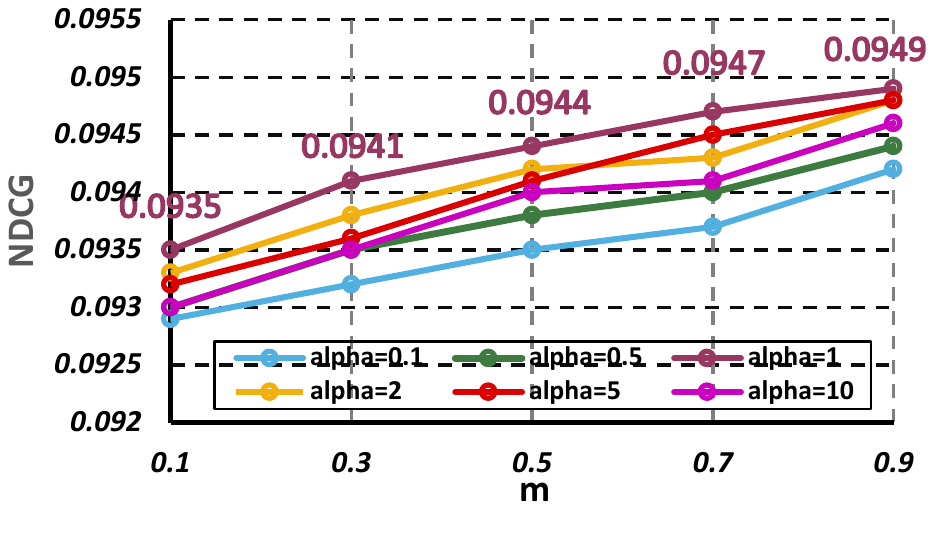}
	\label{fig:figure5_1}
    }
    \subfigure[Recall on LastFM]{
    \includegraphics[width=0.22\textwidth]{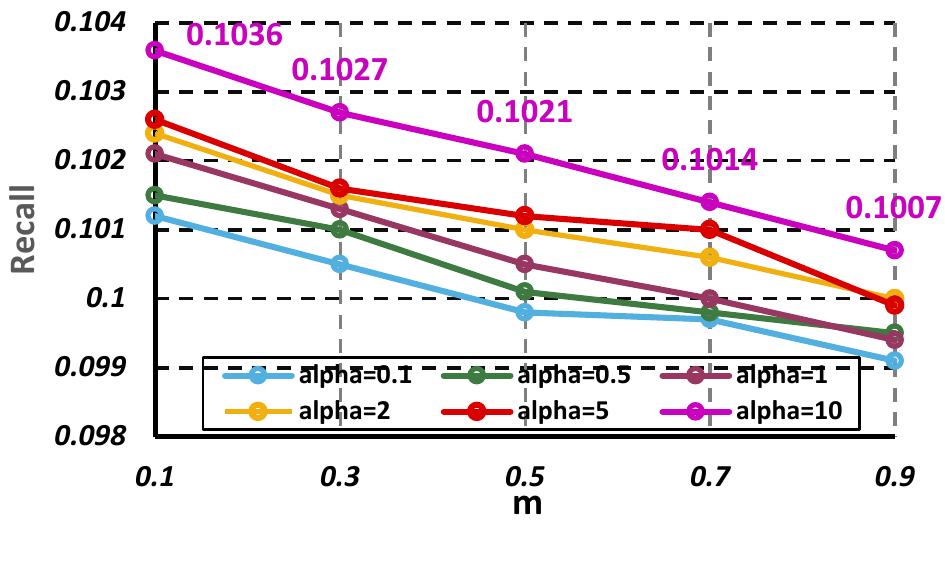}
	\label{fig:figure5_2}
    }
    \subfigure[NDCG on LastFM]{
    \includegraphics[width=0.22\textwidth]{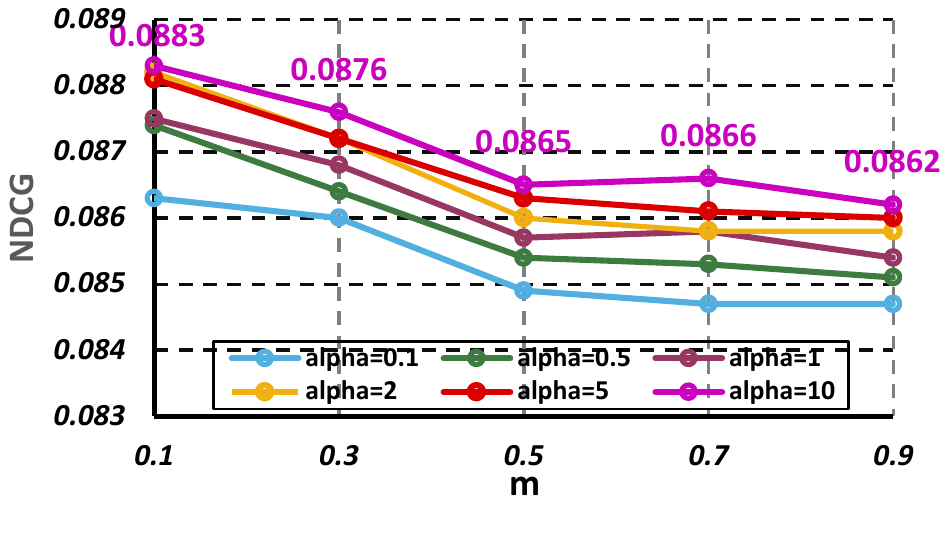}
	\label{fig:figure5_2}
    }
    \subfigure[Recall on Yelp2018]{
    \includegraphics[width=0.22\textwidth]{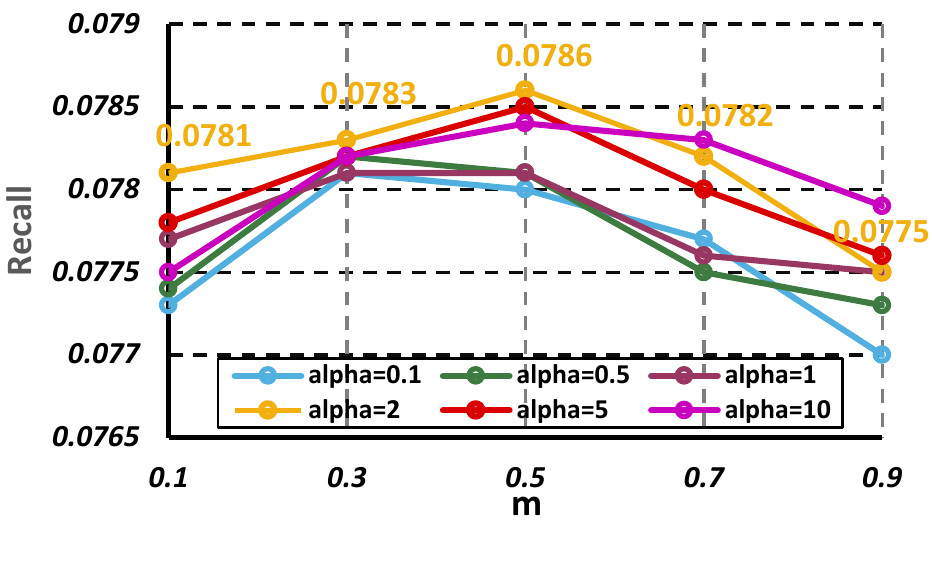}
	\label{fig:figure5_3}
    }
    \subfigure[NDCG on Yelp2018]{
    \includegraphics[width=0.22\textwidth]{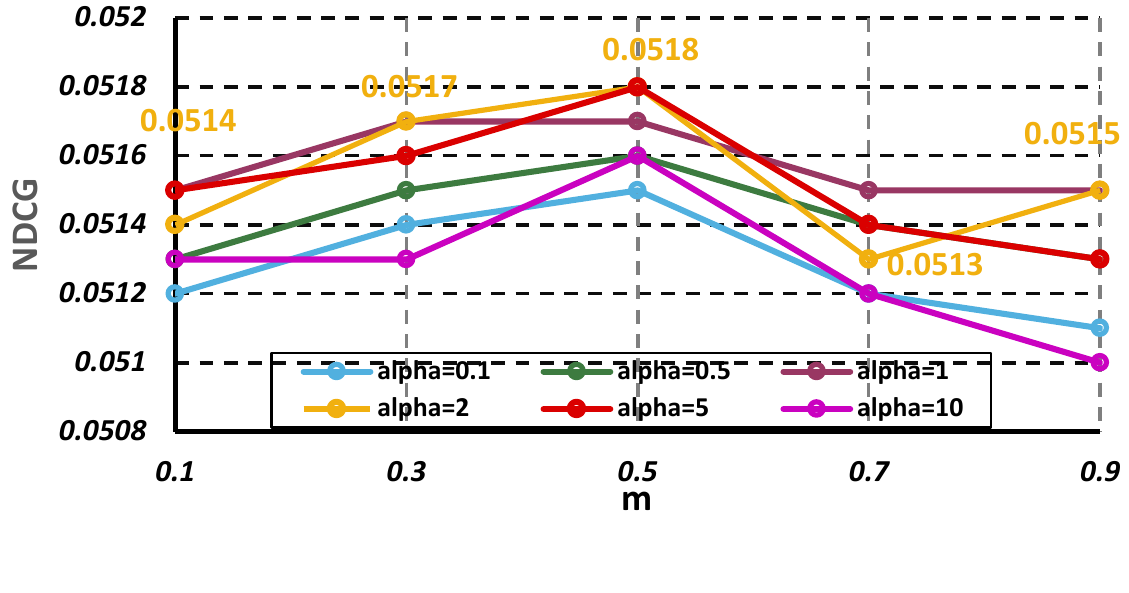}
	\label{fig:figure5_3}
    }
    \caption{Performance of KGCR in terms of margin $m$ and weight $\alpha$ on Amazon-book, LastFM, and Yelp2018. Best viewed in color.}
\end{figure}

\subsection{Case Study}
We randomly selected two users $\#47847$ and $\#50248$ from Amazon-Book, and illustrate their predicted scores of top 20 items in Figure~\ref{fig:figure5}. To be more specific, we decompose the predicted scores into CF scores (\textit{i.e.,} blue bars) and debiased similarity scores (\textit{i.e.,} green bars), and emphasize the ground-true with red rectangles. 

Obviously, the positive items~(\textit{i.e.,} ground-truth samples) achieve higher similarity scores than the others, justifying the debiased similarity scoring again. More importantly, it is easy to find that the debiased similarity scores help the prediction of user-item interactions. In particular, as shown in the left subfigure, we see that four positive items~(\textit{i.e.,} ranking at $5$, $6$, $7$, and $18$) will drop out of the top 20 if we remove the debiased similarity scores. Moreover, in the right subfigure, when forgoing the debiased similarity scores, the ground-truth samples inevitably fall behind in the rankings. These demonstrate the effectiveness of fine-grained user preference on attribute and verify the rationality of our constructed causal graph. 

\subsection{Hyper-parameter Studies}
Studying the object function Eq.~\ref{equ:equ15}, we suggest that two hyper-parameters, \textit{i.e.,} $m$ and $\alpha$, affect the performance of our proposed model. We hence conduct the experiments on three datasets to evaluate them and illustrate the result \textit{w.r.t.} Recall@20 and NDCG@20 in Figure 5. 

Testing KGCR by varying the value of margin $m$ in range of $\{0.1, 0.3, 0.5, 0.7, 0.9\}$, we can see the results \textit{w.r.t.} Recall and NDCG present the different curves on three datasets. It indicates the difference of scores between positive and negative pairs are various. The possible reason is the different number of observed items per user on three datasets. The dense one, like LastFM dataset, is more likely to sample the hard negative item that is similar to the positive item, making their scores closer. 

As for $\alpha$, we test it in the range of $\{0.1, 0.5, 1, 2, 5, 10\}$ and observe that KGCR with the larger value of $\alpha$ probably performs better. We attribute it to our devised loss function $\mathcal{L}_2$. It limits the difference of scores between positive and negative pairs, which may cause the gradient vanishing with the value of $\alpha$ decreases.
\section{Related Work}
\subsection{KG-based Recommendation}
KG-based recommendation roughly falls into three types: embedding-based, path-based, and GCN-based models. 
The embedding-based models~\cite{CKE, DKN, KTUP} mainly focus on learning the knowledge information from the triplets in the KG. Following the assumption of semantic relatedness (\textit{e.g.,} head +relation = tail), they conduct KGE algorithms (\textit{e.g.,} TransE~\cite{TransE} and TransR~\cite{TransR}) on the KG, and then incorporate the learned entity embedding with collaborative signal to improve the recommendation models. For example, CKE~\cite{CKE} adopts TransE algorithm and takes the entity embeddings of items as the context information feeding into the CF-based model. 
Path-based models instead model the patterns of connections among items in the KG to construct the long-range path connecting the user and target item. As one of the representative studies, RippleNet~\cite{Ripplenet} stores the path from the user to historical items and collects these items' representation as the knowledge signal to enrich user representations. Accordingly, it reorganizes the item entities in the KG as the various paths to inject the knowledge information into representations of users. 
More currently, GCN-based models are developed to improve the KG-based recommendation. Typically, they leverage the information propagation mechanism to aggregate the local structure information in the KG. Then, by explicitly or implicitly embedding the learned knowledge into the user and item representations. For instance, KGCN~\cite{KGCN} conducts the graph convolutional operations in an end-to-end manner and captures inter-item relatedness effectively by mining their associated attributes on the KG. 
KGAT combines the KG and user-item bipartite graph as a whole and applies the graph convolutional operations on it. KGIN, a follow-up work, is also built upon the same heterogeneous graph and designs an adaptive aggregation method to learn the fine-grained user intents from the KG.

In summary, these methods either treat the KG as complementary information or expect to construct the item-item correlation via the edge in the KG, where knowledge information merely be used to supplement the collaborative information. Different from the prior studies, we distinguish the knowledge information from the collaborative one and learn the fine-grained user preference on attribute.
\subsection{Causal Inference}
Causal inference have been widely adopted on the visual computing~\cite{yangxun,hanwang}, neural language processing~\cite{fuli1,fuli2}, and information retrieval~\cite{wenjie,DecRS} tasks. The core of causal inference is to abstract the causal-effect relation between the variables from the target task. Based on the causal-effect factor, the spurious relation can be removed from the truly causal relation, so as to empower the robustness and fairness of models. 

In the recommendation domain, Wang~\textit{et al.,}~\cite{DecRS} constructed the causal graph to eliminate the bias caused by user historical interactions in the CF-based recommendation model. Zhang~\textit{et.al.,}~\cite{zhangyang} focused on the popularity bias issue in the recommendation model and used the causal-effect tools to flexibly control the bias for the high-quality prediction. Beyond the CF-based recommendation model, Wang~\textit{et al.,}~\cite{wenjie} mitigated the clickbait issue in the multimedia recommendation model with the counterfactual inference. To resolve the visual bias from the gap between click and purchase behaviors, Qiu~\textit{et al.,}~\cite{Causalrec} developed a new method to retain the supportive visual information and perform visual debiasing.

Different from these methods, we construct the causal graph for the KG-based recommendation model. By exploring the graph, we point out the spurious relations from the causal-effect relation between the variables and develop a new model to eliminate them.
\section{Conclusion and Future Work}
In this work, we propose to model the fine-grained user preference on attribute to improve the recommender system. Towards this end, we construct the causal graph to resolve the challenges in a causal view, which is the first attempt in the KG-based recommendation to the best of our knowledge. 
Analyzing the constructed causal graph, we attribute the challenges into the spurious relation negatively affecting the prediction and develop a Knowledge Graph based Causal Recommendation model~(KGCR) to address the problem. Specifically, we design the deconfounded user preference learning to model the user preference on attribute by removing the confounder between the user preference and her/his interacted attributes. Furthermore, we leverage the counterfactual inference to eliminate the bias misleading the attribute-based similarity scores of user-item pairs.

Despite the promising performance, there remains some problem that should be explored in future work. For instance, the distribution of attributes over all items may cause popularity bias in the interaction prediction. And, how to distinguish the causal-effect relation~\cite{Discovery1,Discovery2} between entities in the KG and leverage it in the recommendation is also a challenging problem. 
\bibliographystyle{IEEEtran}
\bibliography{reference}

\begin{thebibliography}{10}
\providecommand{\url}[1]{#1}
\csname url@samestyle\endcsname
\providecommand{\newblock}{\relax}
\providecommand{\bibinfo}[2]{#2}
\providecommand{\BIBentrySTDinterwordspacing}{\spaceskip=0pt\relax}
\providecommand{\BIBentryALTinterwordstretchfactor}{4}
\providecommand{\BIBentryALTinterwordspacing}{\spaceskip=\fontdimen2\font plus
\BIBentryALTinterwordstretchfactor\fontdimen3\font minus
  \fontdimen4\font\relax}
\providecommand{\BIBforeignlanguage}[2]{{%
\expandafter\ifx\csname l@#1\endcsname\relax
\typeout{** WARNING: IEEEtran.bst: No hyphenation pattern has been}%
\typeout{** loaded for the language `#1'. Using the pattern for}%
\typeout{** the default language instead.}%
\else
\language=\csname l@#1\endcsname
\fi
#2}}
\providecommand{\BIBdecl}{\relax}
\BIBdecl

\bibitem{KGNN-LS}
H.~Wang, F.~Zhang, M.~Zhang, J.~Leskovec, M.~Zhao, W.~Li, and Z.~Wang,
  ``Knowledge-aware graph neural networks with label smoothness regularization
  for recommender systems,'' in \emph{ACM SIGKDD}, 2019, pp. 968--977.

\bibitem{KGAT}
X.~Wang, X.~He, Y.~Cao, M.~Liu, and T.-S. Chua, ``Kgat: Knowledge graph
  attention network for recommendation,'' in \emph{ACM SIGKDD}, 2019, pp.
  950--958.

\bibitem{KGIN}
X.~Wang, T.~Huang, D.~Wang, Y.~Yuan, Z.~Liu, X.~He, and T.-S. Chua, ``Learning
  intents behind interactions with knowledge graph for recommendation,'' in
  \emph{Web Conference}, 2021, pp. 878--887.

\bibitem{KG}
Q.~Wang, Z.~Mao, B.~Wang, and L.~Guo, ``Knowledge graph embedding: A survey of
  approaches and applications,'' \emph{IEEE TKDE}, vol.~29, no.~12, pp.
  2724--2743, 2017.

\bibitem{survey_new}
Z.~Sun, Q.~Guo, J.~Yang, H.~Fang, G.~Guo, J.~Zhang, and R.~Burke, ``Research
  commentary on recommendations with side information: A survey and research
  directions,'' \emph{ECRA}, vol.~37, 2019.

\bibitem{Survey}
Q.~Guo, F.~Zhuang, C.~Qin, H.~Zhu, X.~Xie, H.~Xiong, and Q.~He, ``A survey on
  knowledge graph-based recommender systems,'' \emph{IEEE TKDE}, 2020.

\bibitem{CKE}
F.~Zhang, N.~J. Yuan, D.~Lian, X.~Xie, and W.-Y. Ma, ``Collaborative knowledge
  base embedding for recommender systems,'' in \emph{ACM SIGKDD}, 2016, pp.
  353--362.

\bibitem{DKN}
H.~Wang, F.~Zhang, X.~Xie, and M.~Guo, ``Dkn: Deep knowledge-aware network for
  news recommendation,'' in \emph{Web Conference}, 2018, pp. 1835--1844.

\bibitem{Pathsim}
Y.~Sun, J.~Han, X.~Yan, P.~S. Yu, and T.~Wu, ``Pathsim: Meta path-based top-k
  similarity search in heterogeneous information networks,'' \emph{VLDB},
  vol.~4, no.~11, pp. 992--1003, 2011.

\bibitem{SI2}
B.~Hu, C.~Shi, W.~X. Zhao, and P.~S. Yu, ``Leveraging meta-path based context
  for top-n recommendation with a neural co-attention model,'' in \emph{ACM
  SIGKDD}, 2018, pp. 1531--1540.

\bibitem{KVMN}
J.~Huang, W.~X. Zhao, H.~Dou, J.-R. Wen, and E.~Y. Chang, ``Improving
  sequential recommendation with knowledge-enhanced memory networks,'' in
  \emph{ACM SIGIR}, 2018, pp. 505--514.

\bibitem{DecRS}
W.~Wang, F.~Feng, X.~He, X.~Wang, and T.-S. Chua, ``Deconfounded recommendation
  for alleviating bias amplification,'' in \emph{ACM SIGKDD}, 2021.

\bibitem{causality}
J.~Pearl, \emph{Causality}.\hskip 1em plus 0.5em minus 0.4em\relax Cambridge
  university press, 2009.

\bibitem{MMGCN}
Y.~Wei, X.~Wang, L.~Nie, X.~He, R.~Hong, and T.-S. Chua, ``Mmgcn: Multi-modal
  graph convolution network for personalized recommendation of micro-video,''
  in \emph{ACM MM}, 2019, pp. 1437--1445.

\bibitem{HS-GCN}
H.~Liu, Y.~Wei, J.~Yin, and L.~Nie, ``Hs-gcn: Hamming spatial graph
  convolutional networks for recommendation,'' \emph{IEEE TKDE}, 2022.

\bibitem{MF}
Y.~Koren, R.~Bell, and C.~Volinsky, ``Matrix factorization techniques for
  recommender systems,'' \emph{Computer}, vol.~42, no.~8, pp. 30--37, 2009.

\bibitem{NCF}
X.~He, L.~Liao, H.~Zhang, L.~Nie, X.~Hu, and T.-S. Chua, ``Neural collaborative
  filtering,'' in \emph{Web Conference}, 2017, pp. 173--182.

\bibitem{KGR1}
C.~Wang, M.~Zhang, W.~Ma, Y.~Liu, and S.~Ma, ``Make it a chorus: knowledge-and
  time-aware item modeling for sequential recommendation,'' in \emph{ACM
  SIGIR}, 2020, pp. 109--118.

\bibitem{KGR2}
J.~Gong, S.~Wang, J.~Wang, W.~Feng, H.~Peng, J.~Tang, and P.~S. Yu,
  ``Attentional graph convolutional networks for knowledge concept
  recommendation in moocs in a heterogeneous view,'' in \emph{ACM SIGIR}, 2020,
  pp. 79--88.

\bibitem{KGR3}
T.~Qi, F.~Wu, C.~Wu, and Y.~Huang, ``Personalized news recommendation with
  knowledge-aware interactive matching,'' in \emph{ACM SIGIR}, 2021.

\bibitem{Why}
J.~Pearl and D.~Mackenzie, \emph{The book of why: the new science of cause and
  effect}.\hskip 1em plus 0.5em minus 0.4em\relax Basic books, 2018.

\bibitem{wenjie}
W.~Wang, F.~Feng, X.~He, H.~Zhang, and T.-S. Chua, ``Clicks can be cheating:
  Counterfactual recommendation for mitigating clickbait issue,'' in \emph{ACM
  SIGIR}, 2021, pp. 1288--1297.

\bibitem{yangxun}
X.~Yang, F.~Feng, W.~Ji, M.~Wang, and T.-S. Chua, ``Deconfounded video moment
  retrieval with causal intervention,'' in \emph{ACM SIGIR}, 2021.

\bibitem{NWGM}
K.~Xu, J.~Ba, R.~Kiros, K.~Cho, A.~Courville, R.~Salakhudinov, R.~Zemel, and
  Y.~Bengio, ``Show, attend and tell: Neural image caption generation with
  visual attention,'' in \emph{ICML}, 2015, pp. 2048--2057.

\bibitem{proof1}
S.~Abramovich and L.-E. Persson, ``Some new estimates of the ‘jensen
  gap’,'' \emph{JIA}, vol. 2016, no.~1, pp. 1--9, 2016.

\bibitem{proof2}
X.~Gao, M.~Sitharam, and A.~E. Roitberg, ``Bounds on the jensen gap, and
  implications for mean-concentrated distributions,'' \emph{arXiv preprint
  arXiv:1712.05267}, 2017.

\bibitem{TransE}
A.~Bordes, N.~Usunier, A.~Garcia-Duran, J.~Weston, and O.~Yakhnenko,
  ``Translating embeddings for modeling multi-relational data,'' \emph{NIPS},
  vol.~26, 2013.

\bibitem{IMPGCN}
F.~Liu, Z.~Cheng, L.~Zhu, Z.~Gao, and L.~Nie, ``Interest-aware message-passing
  gcn for recommendation,'' in \emph{Web Conference}.\hskip 1em plus 0.5em
  minus 0.4em\relax ACM, 2021, p. 1296–1305.

\bibitem{cadene2019rubi}
R.~Cadene, C.~Dancette, M.~Cord, D.~Parikh \emph{et~al.}, ``Rubi: Reducing
  unimodal biases for visual question answering,'' \emph{NIPS}, vol.~32, 2019.

\bibitem{BPR}
S.~Rendle, C.~Freudenthaler, Z.~Gantner, and L.~Schmidt-Thieme, ``Bpr: Bayesian
  personalized ranking from implicit feedback,'' \emph{arXiv preprint
  arXiv:1205.2618}, 2012.

\bibitem{GRCN}
Y.~Wei, X.~Wang, L.~Nie, X.~He, and T.-S. Chua, ``Graph-refined convolutional
  network for multimedia recommendation with implicit feedback,'' in \emph{ACM
  MM}, 2020, pp. 3541--3549.

\bibitem{HUIGN}
Y.~Wei, X.~Wang, X.~He, L.~Nie, Y.~Rui, and T.-S. Chua, ``Hierarchical user
  intent graph network for multimedia recommendation,'' \emph{IEEE TMM}, 2021.

\bibitem{DIE}
J.~Pearl, ``Direct and indirect effects,'' \emph{arXiv preprint
  arXiv:1301.2300}, 2013.

\bibitem{BoW}
J.~Pearl and D.~Mackenzie, \emph{The book of why: the new science of cause and
  effect}.\hskip 1em plus 0.5em minus 0.4em\relax Basic books, 2018.

\bibitem{Ripplenet}
H.~Wang, F.~Zhang, J.~Wang, M.~Zhao, W.~Li, X.~Xie, and M.~Guo, ``Ripplenet:
  Propagating user preferences on the knowledge graph for recommender
  systems,'' in \emph{CIKM}, 2018, pp. 417--426.

\bibitem{KTUP}
Y.~Cao, X.~Wang, X.~He, Z.~Hu, and T.-S. Chua, ``Unifying knowledge graph
  learning and recommendation: Towards a better understanding of user
  preferences,'' in \emph{Web Conference}, 2019, pp. 151--161.

\bibitem{MKR}
H.~Wang, F.~Zhang, M.~Zhao, W.~Li, X.~Xie, and M.~Guo, ``Multi-task feature
  learning for knowledge graph enhanced recommendation,'' in \emph{Web
  Conference}, 2019, pp. 2000--2010.

\bibitem{CKAN}
Z.~Wang, G.~Lin, H.~Tan, Q.~Chen, and X.~Liu, ``Ckan: Collaborative
  knowledge-aware attentive network for recommender systems,'' in \emph{ACM
  SIGIR}, 2020, pp. 219--228.

\bibitem{Xavier}
X.~Glorot and Y.~Bengio, ``Understanding the difficulty of training deep
  feedforward neural networks,'' in \emph{JMLR}, 2010, pp. 249--256.

\bibitem{Adam}
D.~P. Kingma and J.~Ba, ``Adam: A method for stochastic optimization,'' in
  \emph{ICML}, 2015, pp. 1--16.

\bibitem{LightGCN}
X.~He, K.~Deng, X.~Wang, Y.~Li, Y.~Zhang, and M.~Wang, ``Lightgcn: Simplifying
  and powering graph convolution network for recommendation,'' in \emph{ACM
  SIGIR}, 2020, pp. 639--648.

\bibitem{TransR}
Y.~Lin, Z.~Liu, M.~Sun, Y.~Liu, and X.~Zhu, ``Learning entity and relation
  embeddings for knowledge graph completion,'' in \emph{AAAI}, 2015.

\bibitem{KGCN}
H.~Wang, M.~Zhao, X.~Xie, W.~Li, and M.~Guo, ``Knowledge graph convolutional
  networks for recommender systems,'' in \emph{Web Conference}, 2019, pp.
  3307--3313.

\bibitem{hanwang}
T.~Wang, J.~Huang, H.~Zhang, and Q.~Sun, ``Visual commonsense r-cnn,'' in
  \emph{CVPR}, 2020, pp. 10\,760--10\,770.

\bibitem{fuli1}
F.~Feng, J.~Zhang, X.~He, H.~Zhang, and T.-S. Chua, ``Empowering language
  understanding with counterfactual reasoning,'' \emph{arXiv preprint
  arXiv:2106.03046}, 2021.

\bibitem{fuli2}
F.~Feng, W.~Huang, X.~He, X.~Xin, Q.~Wang, and T.-S. Chua, ``Should graph
  convolution trust neighbors? a simple causal inference method,'' in \emph{ACM
  SIGIR}, 2021, pp. 1208--1218.

\bibitem{zhangyang}
Y.~Zhang, F.~Feng, X.~He, T.~Wei, C.~Song, G.~Ling, and Y.~Zhang, ``Causal
  intervention for leveraging popularity bias in recommendation,'' in \emph{ACM
  SIGIR}, 2021.

\bibitem{Causalrec}
R.~Qiu, S.~Wang, Z.~Chen, H.~Yin, and Z.~Huang, ``Causalrec: Causal inference
  for visual debiasing in visually-aware recommendation,'' in \emph{ACM MM},
  2021, pp. 3844--3852.

\bibitem{Discovery1}
S.~Zhu, I.~Ng, and Z.~Chen, ``Causal discovery with reinforcement learning,''
  \emph{arXiv preprint arXiv:1906.04477}, 2019.

\bibitem{Discovery2}
Y.~Wang, V.~Menkovski, H.~Wang, X.~Du, and M.~Pechenizkiy, ``Causal discovery
  from incomplete data: a deep learning approach,'' \emph{arXiv preprint
  arXiv:2001.05343}, 2020.

\end{thebibliography}
\end{document}